\documentclass[10pt]{article}
\usepackage{amsmath,hyperref}
\usepackage{amssymb,amsthm}
\usepackage{amsfonts}
\usepackage{subcaption}
\usepackage{graphicx}
\graphicspath{{figs/}}
\usepackage[left=2.5cm,right=2.2cm,top=0.5cm,bottom=1.3cm,includeheadfoot]{geometry}
\usepackage{indentfirst}
\usepackage{color}
\usepackage{cite}
\usepackage[notlot,notlof,nottoc]{tocbibind}

\begin{document}
	\begin{center}
		{\Large\bf The chaotic behavior of the Bianchi IX model under the influence\\\vskip 2mm of quantum effects}
		\vskip 5mm
		{\large
			Martin Bojowald$^1$, David Brizuela$^2$, Paula Calizaya Cabrera$^3$, and Sara F. Uria$^2$}
		\vskip 3mm
		{\sl $^1$Institute for Gravitation and the Cosmos,
			The Pennsylvania State University, 104 Davey Lab,\\ University Park, PA 16802, USA}\\
		{\sl $^2$Department of Physics and EHU Quantum Center, University of the Basque Country UPV/EHU,\\
			Barrio Sarriena s/n, 48940 Leioa, Spain}\\
		{\sl $^3$	
			Department of Physics and Astronomy, Louisiana State University, Baton Rouge, LA 70803, USA}
	\end{center}
	\begin{abstract}
          A quantum analysis of the vacuum Bianchi IX model is performed,
          focusing in particular on the chaotic nature of the system.
          The framework constructed here is general enough for the
            results to apply in the context of any theory of quantum gravity,
            since it includes only minimal approximations that make it possible to
            encode the information of all quantum degrees of freedom in the
            fluctuations of the usual anisotropy parameters.  These
          fluctuations are described as canonical variables that extend the
          classical phase space. In this way, standard methods for dynamical
          systems can be applied to study the chaos of the model. Two specific
          methods are applied that are suitable for
            time-reparameterization invariant systems. First, a generalized
          version of the Misner-Chitre variables is constructed, which
          provides an isomorphism between the quantum Bianchi IX dynamics and
          the geodesic flow on a suitable Riemannian manifold,
          extending, in this way, the usual billiard picture. Secondly, the fractal dimension of the boundary
          between points with different outcomes in the space of initial data
          is numerically analyzed. While the quantum system remains
            chaotic, the main conclusion is that its strength is considerably
          diminished by quantum effects as compared to its classical
          counterpart.
	\end{abstract}

	\section{Introduction}\label{sec:intro}
	
	According to the Belinski-Khalatnikov-Lifshitz (BKL) conjecture
        \cite{BKL} the dynamics of the spacetime near a generic spacelike
        singularity becomes local (in the sense that different spatial points decouple), oscillatory, and
	dominated by pure gravity.
	Therefore, the vacuum Bianchi IX (also known as Mixmaster) model \cite{Mixmaster_I,Mixmaster}
	is assumed to provide a 
	proper characterization of the evolution of each of these points.
	A formal proof of the BKL conjecture is not yet available,
	but it is supported by a large number of numerical studies \cite{Berger,Garfinkle,Heinzle}.
	Therefore, the general working assumption is that this conjecture is true,
	and thus the study of the Mixmaster model has received much attention.
	In particular, regarding its specific properties, there is a wide literature
	on the chaotic nature of the Mixmaster---and hence the BKL---dynamics.
	In fact, in the context of chaos, this is one of the most discussed solutions of the Einstein equations \cite{Deterministic_chaos_book, Barrow_1}. However,
	due to the nature of general relativity, usual dynamical methods to study chaos cannot be applied, as they are
	not covariant (observer independent). Hence, different techniques have
        been proposed and implemented to prove the chaotic nature of the Mixmaster model.
	
	The first attempts in this direction were already made by the original authors of the BKL conjecture \cite{BKL}. Specifically, they showed that, under certain assumptions,
	the full dynamics of the system given by the Einstein equations can be approximated
	by a discrete map. A large number of iterations of this map corresponds thus to the asymptotic limit of the evolution
	towards the singularity. They were then able to prove that the dependence on the initial conditions vanished
	in such a limit. A few years later, making use of this same discrete map, but considering
	the topological and metric entropies of its related one-dimensional Poincar\'e section,
	similar conclusions were found in Refs.~\cite{Barrow_2,Barrow_3,Barrow_4}.
	Thus, both approaches proved that the discrete dynamics is chaotic, though the full dynamics was still
	to be analyzed.
	
	Subsequent studies were then devoted to an analysis of the full
        dynamics, solving numerically the corresponding Einstein
        equations. Most of them focused on the computation of the
        Lyapunov exponents, since, in usual dynamical systems, a positive
        value of any of these exponents characterizes the system to be
        chaotic. However, in the present context, this led to several
        (apparent) contradictions: depending on the time parameterization, both
        positive and nonpositive values were obtained
        \cite{Rugh,Francisco,Berger_numerical,Hobill, Ferraz,Szydlowski}.  By
        construction, all these results were obtained in a given coordinate
        system and were thus observer dependent. Therefore, at the time, the
        usual explanation of the controversy pointed to a violation of
        general covariance.  However, this was not the case: as Motter
        later showed \cite{Motter_lyap}, the sign of the Lyapunov exponent, if
        properly defined, is invariant under coordinate transformations.  In
        fact, another important conclusion of this work was that none of the
        mentioned investigations satisfied the requirements for a proper
        definition of the Lyapunov exponent.  More precisely, in order to have
        an invariant meaning, the computation of this exponent must be done
        in certain set of coordinates, so that the system meets some specific
        conditions\cite{Motter_lyap}.  Hence, one of these suitable coordinate
        systems was chosen in \cite{Montani_paper} and, just by an analytical
        computation of the exponent, it was confirmed that the full Mixmaster
        dynamics is indeed chaotic. More specifically, it was proven that it
        is isomorphic to a billiard on the two-dimensional Lobachevsky plane,
        which is well known to exhibit chaos, as it has convex and therefore
        defocusing walls \cite{Math_concave}.
	
	On the other hand, another completely different approach was followed
        to test the chaotic nature of the Mixmaster dynamics, by means of
        observer independent---and therefore covariant---fractal methods
        \cite{Cornish_Levin,Motter_fract}. These studies considered both the
        discrete map, as well as the full dynamics, and showed that this
        system is characterized by the presence of a strange repeller with
        fractal dimensions, which ensures chaos. Moreover, the authors
          of Ref.~\cite{Cornish_Levin} were able to quantify chaos by
        computing different relevant quantities, such as the uncertainty exponent or the multifractal dimensions of the repeller. 
	
	However, even if it is now generally accepted that the classical Mixmaster dynamics is chaotic,
	quantum effects are expected to be relevant asymptotically, and they might completely modify
	the classical picture in the extreme regimes close to the spacelike singularity.
	Several analyses can be found in the literature, proposing different quantizations of this model,
	and focused on a variety of questions, such as the avoidance of the singularity, or the modifications of its oscillatory dynamics (see, e.g., Refs. \cite{Berger_quantum,Bojowald_Date_1,Vaulin,Benini_Montani,Wilson-Ewing-1,Ashtekar,Bergeron,Bae,Bergeron_2,Garfinkle_quantum,Corichi,Kiefer,Wilson-Ewing_2,Piechocki,Montani_polymer,Quantum_Bianchi_IX}). In particular,
	some works have been devoted to analyzing the survival of chaos, but so far with mixed results \cite{Coley,Bojowald_Date_2,Bergeron_3,Montani_chaos_polymer}. In these approaches specific properties
	that may reduce chaos have been identified (such as, for instance, an isotropization of the system,
	or a possible bounded curvature), but none of them has studied these effects in a general way,
	such that it can be applied to any quantum-gravity model.

        Therefore, in the present paper,
	we present a systematic semiclassical analysis that will allow us to
	study how quantum effects modify the chaotic behavior of the Mixmaster model.
	Our framework is based on a decomposition of the wave function
	into its infinite set of moments, and we will introduce certain assumptions
	(a Gaussian-like state and small fluctuations) so that quantum effects will be
	encoded in a finite set of canonical variables. In this way, the system can
	be described in a phase space, which constitutes an extension of the classical phase space by
	the fluctuation degrees of freedom, and  thus usual dynamical-systems techniques
	can be applied to study chaos. In particular, the two methods mentioned
	above (the computation of the covariant Lyapunov exponent and fractal methods)
	will be applied to this quantum system. Our main conclusion will be that, even
	if the quantum system is still chaotic, quantum effects significantly reduce the level of chaos.
	
	The paper is organized as follows. In Sec.~\ref{sec:classical_model}
        the classical Mixmaster model is
        presented. Sec.~\ref{sec:quantum_model} describes the construction of
        the framework that will describe the Mixmaster dynamics in a
        semiclassical regime. Sec.~\ref{sec:analysis_quantum} contains a
        qualitative description of this dynamics. In
        Sec.~\ref{sec:analysis_chaos} the chaotic nature of the model
        is studied following two different approaches. More precisely, in
        Sec.~\ref{sec:Lyapunov} the covariant Lyapunov exponent is computed,
        while in Sec.~\ref{sec:fractal} the fractal dimension of the repeller
        set in the space of initial data is analyzed.  Finally, in
        Sec.~\ref{sec:discussion} the main results of the paper are discussed
        and summarized.
	
	\section{Classical model}\label{sec:classical_model}
	
		\begin{figure}
		\centering
		\includegraphics[width=0.55\linewidth]{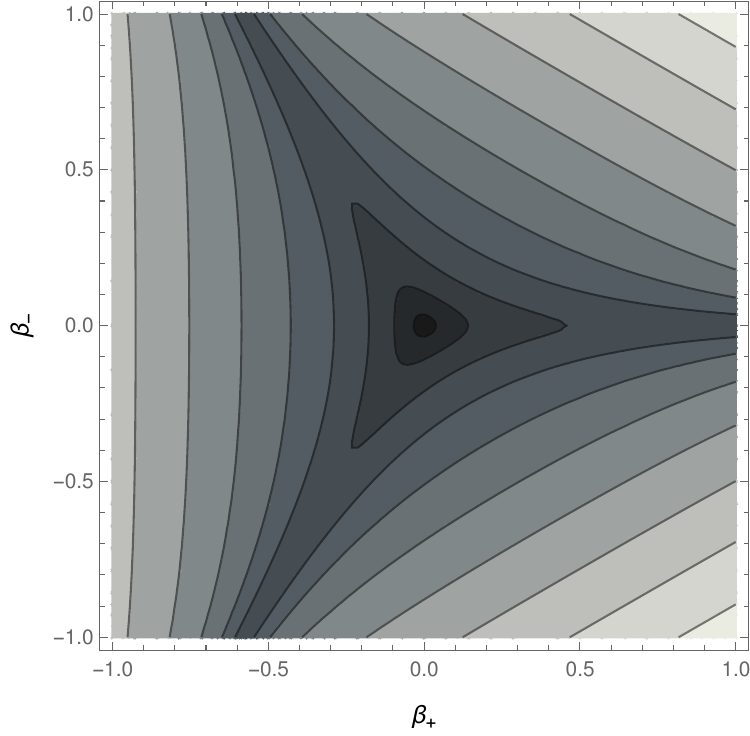}
		\caption{Contour plot of the classical Bianchi IX potential $V(\beta_+,\beta_-)$ given by Eq. \eqref{potential_Bianchi_IX}.}
		\label{fig:equipotential}
	\end{figure}
	
	The vacuum Bianchi IX model, also known as the Mixmaster model,
	is described by the metric
	\begin{align}\label{metric}
	ds^2=-N^2 dt^2+\sum_{i=1}^3a_ {i}^2\sigma_i^2,
	\end{align}
	where $N$ is the lapse function, $a_i$ are the expansion rates for each of the three space
	directions, and $\sigma_i$ are 1-forms on the three sphere that satisfy
	$d\sigma_i=\frac{1}{2}\epsilon_{ijk}\sigma_j\wedge \sigma_k$. A particular
	parameterization of these forms can be given in terms of the Euler angles, 
	\begin{align}\label{dual_forms}
	\begin{split}	&\sigma_1:=\sin\psi
	d\theta-\cos\psi\sin\theta d\phi,
	\\
	&\sigma_2:=\cos\psi d\theta+\sin\psi\sin\theta d\phi,
	\\
	&\sigma_3:=-d\psi-\cos\theta d\phi.
	\end{split}
	\end{align}
	As first pointed out by Misner\cite{Mixmaster}, it is convenient
	to define basic variables that, instead of describing the change of every
	individual spatial direction, determine the change of the volume
	and shape of the three-dimensional geometry. More precisely, one can perform a change of variables from the
	three expansion rates $a_i$ to a new set of variables given by $\alpha$,
	defined as
	\begin{equation}
	e^{\alpha}=(a_1a_2a_3)^{1/3},
	\end{equation}
	encoding the volume of the spatial sections, and the two shape parameters
	$\beta_+$ and $\beta_-$, defined as the logarithmic rates
	\begin{align}\label{misner_variables}
	\begin{split}
	\beta_{+}&:=-\frac{1}{2}\ln \left[{\frac{a_3}{(a_1a_2a_3)^{1/3}}}\right],
	\\
	\beta_{-}&:=\frac{1}{2\sqrt{3}}\ln \left({\frac{a_1}{a_2}}\right),
	\end{split}
	\end{align}
	as a description of the anisotropy.
	
	In terms of these so-called Misner variables, the Hamiltonian constraint for the Mixmaster model reads
	\begin{equation}\label{constraint}
	{\cal C}=\frac{1}{2} e^{-3 \alpha}
	\left(-p_\alpha^2+p_{-}^2+p_{+}^2\right)+e^{\alpha} V(\beta_+,\beta_-)=0.
	\end{equation}
	This constraint equation requires an energy balance between the potential
	\begin{align}\label{potential_Bianchi_IX}
	&V(\beta_+,\beta_-):=\frac{1}{6}\left[
	e^{-8\beta_+}+2e^{4\beta_+}\left(
	\cosh(4\sqrt{3}\beta_-)-1
	\right)-4e^{-2\beta_+}\cosh(2\sqrt{3}\beta_-)
	\right],
	\end{align}
	whose contour plot is shown in Fig.~\ref{fig:equipotential},
	and the kinetic term, which is a quadratic combination of the momenta
	$p_{\alpha}:=-(e^{3\alpha}/N) d\alpha/dt$ and $p_{\pm}:=(e^{3\alpha}/N) d\beta_{\pm}/dt$.
	
	The dynamics of this model describes an expanding (or contracting) universe. Therefore the volume, and thus the variable $\alpha$,
	are monotonic functions of coordinate time, which allows us to define $\alpha$ as the internal
	time variable by imposing the gauge $\alpha=t$. The deparameterization of the above constraint
	then leads to the physical (generically nonvanishing) Hamiltonian,
	\begin{align}\label{hamiltonian}
	H:=&-p_\alpha=\big[p_+^2+p_-^2+2e^{4\alpha}V(\beta_+,\beta_-)\big]^{1/2}.
	\end{align}
	In this gauge, the independent dynamical variables are the shape parameters $\beta_\pm$
	along with their momenta $p_\pm$. Their equations of motion are obtained by
	computing their Poisson brackets with the Hamiltonian,
	\begin{align}
	\label{equations_motion_classical_beta_+}
	&\dot{\beta}_{+}=
	\{
	\beta_+,H
	\}=\frac{p_{+}}{H},
	\\
	\label{equations_motion_classical_beta_-}
	&\dot{\beta}_{-}=
	\{
	\beta_-,H
	\}=\frac{p_{-}}{H},
	\\
	&
	\label{equations_motion_classical_p_+}
	\dot{p}_{+}=\{
	p_+,H
	\}=-\frac{e^{4\alpha}}{H}
	\frac{\partial V(\beta_+,\beta_-)}{\partial \beta_+},
	\\
	\label{equations_motion_classical_p_-}
	&
	\dot{p}_{-}=\{
	p_-,H
	\}=-\frac{e^{4\alpha}}{H}
	\frac{\partial V(\beta_+,\beta_-)}{\partial \beta_-},
	\end{align}
	where the dot stands for a derivative with respect to $\alpha$.
	
	\section{Quantum model}\label{sec:quantum_model}
	
	We will perform a canonical quantization of the system by promoting the classical
	Hamiltonian \eqref{hamiltonian} to a quantum operator $\hat{H}$, which will define
	the dynamics of the wave function by the usual Schr\"odinger equation
	$\hat{H}\Psi=i\hbar \partial\Psi/\partial\alpha$. However, instead
	of analyzing the evolution of the wave function, we will directly study the dynamics of the
	central moments,
	\begin{equation}\label{def_moments}
	\Delta(\beta_+^i\beta_-^jp_+^kp_-^l):=\langle
	(\hat{\beta}_+-\beta_+)^i(\hat{\beta}_--\beta_-)^j(\hat{p}_+-p_+)^k
	(\hat{p}_--p_-)^l
	\rangle_{\text{Weyl}},
	\end{equation} 
	where the subscript Weyl implies a completely symmetric ordering of
        the basic operators, the indices $i$, $j$, $k$, and $l$ are
        nonnegative integers, and, from this point on, we define
        $\beta_+:=\langle\hat\beta_+\rangle$,
        $\beta_-:=\langle\hat\beta_-\rangle$, $p_+:=\langle \hat p_+\rangle$,
        and $p_-:=\langle \hat p_-\rangle$.  For such a purpose, following the
        framework presented in \cite{moments}, one needs to define an
        effective Hamiltonian $H_Q$ as the expectation value of the
        Hamiltonian operator $H_Q:=\langle\hat H\rangle$. By performing a
        formal series expansion, $H_Q$ can be written in terms of the
        expectation values of the basic variables, $\langle\beta_+\rangle$,
        $\langle\beta_-\rangle$, $\langle p_+\rangle$, $\langle p_-\rangle$,
        and the central moments defined above.  The time derivative of each of
        these variables is then given by its Poisson bracket with the
        effective Hamiltonian $H_Q$, which can be computed making use of the
        definition
\begin{equation}\label{Poisson}
        \{\langle \hat X \rangle, \langle \hat Y \rangle\}=
        \frac{1}{i\hbar}\langle[\hat X,\hat Y]\rangle
      \end{equation}
      for any two operators
        $\hat X$ and $\hat Y$, and extended to products of expectation values
        by using the Leibniz rule. Explicit formulas for the Poisson brackets
        of moments can be found in  \cite{moments,moments_2011}.

        In this way, instead of dealing with a partial differential equation
        for the wave function, one needs to solve a system of ordinary
        differential equations.  Both types of differential equations
        describe the very same dynamics  and solving ordinary
          differential equations is usually an easier task. However, in
          the present case it does pose a challenge as the infinitely many
          moments \eqref{def_moments} are evolved by a large number of
          equations that in general are strongly coupled for nonquadratic
          Hamiltonians. For practical applications, some approximation is
        therefore necessary in order to deal with this system. A standard
        approach is to introduce a truncation by neglecting all moments higher
        than a certain order, which leads to a finite system that can be
        solved numerically.  In this context the order of a given moment
        \eqref{def_moments} is defined as the sum of its indices
        $i+j+k+l$. However, in the present paper, we will follow another
        approach by imposing two different assumptions, spelled out in
          the next subsection. Instead of truncating the infinite tower of
        moments by setting infinitely many moments equal to zero, the
        new assumptions implement a closure condition that maintains
          nonvanishing higher moments but assumes suitable expressions for
          them in a finite-dimensional parameterization. The specific closure
          assumption will allow us to obtain an exact summation of the infinite
        series that defines the effective Hamiltonian $H_Q$, in such a way
        that we will be able to work with an extended though
        finite-dimensional phase space.

\subsection{Closure conditions}
          
Our first approximation concerns the definition of the effective Hamiltonian
$H_Q$.  As mentioned above, this expression is defined as the
  expectation value of the Hamiltonian operator $\hat{H}$, such that its Hamilton's
  equation with the Poisson bracket (\ref{Poisson}) are equivalent to the
  evolution of moments of a state following the Schr\"odinger dynamics. The
  resulting $H_Q$ therefore depends on the ordering chosen for the operator
  $\hat{H}$. For instance, if $\hat{H}$ is Weyl-ordered, the corresponding
  $H_Q$ directly depends on the moments defined with the same ordering through
  a Taylor expansion, and there are no additional terms with an explicit
  dependence on $\hbar$ from re-ordering products.

   In the present case, it is more convenient to assume that
         $\hat{H}^2$ is Weyl-ordered and to define the quantum Hamiltonian as
         $H_Q:=\sqrt{\langle\hat{H}^2\rangle}$, an assumption that can also be
         motivated by first quantizing the constraint (\ref{constraint}) and
         then deparameterizing after quantization. The evolution generated by
           $H_Q$ does not exactly correspond to the Schr\"odinger evolution
           generated by $\hat{H}$. However,
	these two possible definitions differ only by a fluctuation term
        because we have the exact relation
	\begin{align}\label{approx_H}
	\langle\hat{H}^2\rangle=\langle\hat{H}\rangle^2
	+\langle(
	\hat{H}-\langle{\hat{H}}\rangle
	)^2\rangle.
	\end{align}
	Therefore, as long as the relative fluctuations
        $\langle(\hat{H}-\langle{\hat{H}}\rangle)^2\rangle/H^2$ are small, the
        effective dynamics given by $H_Q$ will accurately describe the Schr\"odinger flow of states.
        
	The main reason to perform this approximation is that it
          implies a greatly simplified expression for the effective Hamiltonian in
        terms of the moments.  More precisely, performing an expansion around
        the expectation values of the basic operators, the square of the
        effective Hamiltonian reads
	\begin{align}\label{quantum_hamiltonian}
	H_Q^2&=\langle\hat{H}^2(\hat{\beta}_+,\hat{\beta}_-,
	\hat{p}_+,\hat{p}_-)\rangle
	=\sum_{i,j,k,l=0}^{\infty}\frac{1}{i!j!k!l!}
	\frac{\partial^{i+j+k+l} H^2(\beta_+,\beta_-,p_+,p_-)}{\partial \beta_+^{i}\partial \beta_-^{j}p_+^kp_-^l}
	\Delta(\beta_+^i\beta_-^jp_+^kp_-^l)
	\\\nonumber
	&=p_+^2+\Delta(p_+^2)+p_-^2+\Delta(p_-^2)+
	e^{4\alpha}\sum_{i,j=0}^{\infty}\frac{1}{i!j!}
	\frac{\partial^{i+j} V(\beta_+,\beta_-)}{\partial \beta_+^{i}\partial \beta_-^{j}}
	\Delta(\beta_+^i\beta_-^j),
	\end{align}
	where, as already mentioned, we have defined $\hat{H}^2$ to be
        Weyl-ordered. Note that $H$ is the classical Hamiltonian and, in the
        last equality, its precise form \eqref{hamiltonian} has been taken
        into account.  In particular, since $H^2$ is quadratic in the momenta
        $p_\pm$, only their fluctuations appear in this effective Hamiltonian,
        and higher-order moments only depend on the shape parameters. One can
        already see at this point that the structure of this effective
        Hamiltonian is quite similar to the classical one: it has a kinetic
        part, which is, in some sense, quadratic in momentum variables, and a
        potential term, that, up to the global factor $e^{4\alpha}$, only
        depends on the position variables. We will make this structure more
        explicit below.
	
	At this point one could obtain the equations of motion for the expectation values of basic operators
	and the moments by computing their Poisson brackets with $H_Q$. However, the equations are
	very complicated, in particular because the moments do not have canonical brackets with one another.
	Nevertheless, according to Darboux' theorem, and its extension to nonsymplectic manifolds, any Poisson manifold can be
	described locally by canonical coordinates, given by generalized positions and momenta, as well
	as Casimir variables. For instance, for fluctuations and correlations one can define the following
	change of variables
	\cite{JK,ABCL,P,VZ,Martin_moments},
	\begin{align}\label{second_moments}
	\begin{split}
	&\Delta(p_{+}^2)=p_{s_1}^2+\frac{U_1}{s_1^2},\quad\Delta(\beta_+^2)=s_1^2,
	\\
	&\Delta(p_-^2)=p_{s_2}^2+\frac{U_2}{s_2^2},\quad\Delta(\beta_-^2)=s_2^2,
	\\
	&\Delta(\beta_+p_+)=s_1p_{s_1},
	\quad\Delta(\beta_-p_-)=s_2p_{s_2},
	\end{split}
	\end{align}
      ignoring cross-correlations between the two pairs of degrees of freedom.
	From these definitions, we note that only the sign of either $s_i$ or $p_{s_i}$
	has physical meaning: both $(s_i,p_{s_i})$ and $(-s_i,-p_{s_i})$ represent the same physical state.
	Thus, we will choose $s_i$ to be positive definite (since it parameterizes the fluctuation
	of $\beta_\pm$, it cannot be vanishing), and let $p_{s_i}$ be defined on the whole real line.
	Under a second-order truncation (neglecting all moments of an order three and higher),
	$s_{i}$ and $p_{s_i}$ are canonically conjugate, satisfying $\{s_i,p_{s_j}\}=\delta_{ij}$, for $i,j=1,2$,
	while $U_i$ are Casimir variables that have vanishing Poisson brackets with the different variables.
	These encode the information about the Heisenberg uncertainty relation, since they obey
	\begin{align}\label{uncert_heis}
	\begin{split}
	\Delta(\beta_+^2)\Delta(p_+^2)-\Delta(\beta_+p_+)^2=U_1\geq\frac{\hbar^2}{4},
	\\
	\Delta(\beta_-^2)\Delta(p_-^2)-\Delta(\beta_-p_-)^2=U_2\geq\frac{\hbar^2}{4}.
	\end{split}
	\end{align}
	
	For higher-order moments, the corresponding canonical variables are
        hard to find explicitly, and their expressions are usually lengthy in
        cases in which they are known (see Ref.~\cite{Martin_moments} for a
        fourth-order derivation for a single pair of classical degrees of
        freedom, as well as a second-order mapping for two pairs of
          degrees of freedom with cross-correlations). Since
        $\Delta(\beta_+^i\beta_-^j)$ are the only higher-order moments that
        appear in the effective Hamiltonian \eqref{quantum_hamiltonian}, the
        second approximation we will perform here is to parameterize these
        higher-order moments solely in terms of the $s_i$ defined above. In order to
        choose a meaningful parameterization, we base our choice in the form of
        the moments for a Gaussian state and impose
	\begin{align}\label{approx_gaussian_moments}
	\Delta(\beta_+^{2n}\beta_-^{2m})=\frac{s_1^{2n}s_2^{2m}}
	{2^n2^m} \frac{(2n)!(2m)!}{n!m!}
	\end{align}
	for nonnegative integers $n,m$, while
        $\Delta(\beta_+^{i}\beta_-^{j})=0$ otherwise.  With this second
        approximation we are thus assuming that the moments corresponding to
        the shape parameters keep the Gaussian form all along the
        evolution, an assumption that can
        be considered valid in a semiclassical context.
        More precisely, since, as will be explained in Sec. \ref{sec:analysis_quantum},
        the classical dynamics follows periods of free evolution separated from one another by quick reflections on steep potential walls, a semiclassical state can be expected to maintain nearly Gaussian form for a considerable amount of time. In particular, the state is not expected to split up into separate wave packets and develop multimodality, as it would happen for instance in tunneling situations. Our assumptions are therefore justified and they may be tested further by including higher-order terms in the Hamiltonian, which we leave for future work.

\subsection{Summation}
        
	Replacing now the form of the moments \eqref{approx_gaussian_moments} in the effective Hamiltonian 
	\eqref{approx_H}, it is possible to perform exactly the infinite sum. In this way, 
	we obtain the following closed form for the effective Hamiltonian,
	\begin{align}\label{quantum_hamiltonian_approx}
	H_Q^2=\langle\hat{H}^2\rangle&=
	p_+^2+p_-^2+p_{s_1}^2+p_{s_2}^2+\frac{U_1}{s_1^2}+\frac{U_2}{s_2^2}
	+2e^{4\alpha}V_{Q}(\beta_+,\beta_-,s_1, s_2),
	\end{align}
	with the corresponding effective potential defined as
	\begin{align}\label{quantum_potential}
	\hspace{-0.4cm}V_{Q}(\beta_+,\beta_-,s_1, s_2):=
	\frac{1}{6}
	\bigg[
	e^{-8\beta_++32s_1^2}
	+2
	e^{4\beta_++8s_1^2}
	\left(
	e^{24s_2^2}
	\cosh(4\sqrt{3}\beta_-)
	-1
	\right)
	-4
	e^{-2\beta_++2s_1^2+6s_2^2}\cosh(2\sqrt{3}\beta_-)
	\bigg].
	\end{align}
	Hence, the only relevant variables of our quantum model will be $\beta_{\pm}$, $p_{\pm}$, $s_i$,
	$p_{s_i}$, and $U_i$ with $i=1,2$. In fact, the variables $U_i$ are constants of motion. In the Hamiltonian
	they appear as centrifugal potential terms and thus can be regarded as an angular momentum that prevents
	the position variables $s_i$ from reaching the origin, enforcing, in this way, the uncertainty relation.
	The Hamiltonian has now a clear structure of a kinetic part, quadratic in momenta, plus a
	potential term, which, up to the global factor $e^{4\alpha}$, only depends on position variables.
	The classical phase space, given by the two couples $(\beta_\pm,p_\pm)$, is thus
	enlarged by the two new degrees of freedom described by $(s_i,p_{s_i})$ and by the parameters
	$U_i$, which encode
	the quantum effects. The vanishing of these quantum variables leads to the classical limit,
	as can be easily seen by comparing \eqref{quantum_hamiltonian_approx} with its classical
	counterpart \eqref{hamiltonian}.
	
	The equations of motion for the different variables read
	\begin{align}
	\label{equations_motion_quantum_beta}
	&\dot{\beta}_{\pm}=
	\{
	\beta_\pm,H_Q
	\}=\frac{p_{\pm}}{H_Q},
	\\
	&
	\label{equations_motion_quantum_p}
	\dot{p}_{\pm}=\{
	p_\pm,H_Q
	\}=-\frac{e^{4\alpha}}{H_Q}
	\frac{\partial V_Q}{\partial \beta_\pm},\\
	&
	\label{equations_motion_quantum_s}
	\dot{s}_{i}=
	\{
	s_i,H_Q
	\}=\frac{p_{s_i}}{H_Q},
	\\
	\label{equations_motion_quantum_ps}
	&
	\dot{p}_{s_i}=\{
	p_{s_i},H_Q
	\}=\frac{1}{H_Q}\left(
	\frac{U_i}{s_i^3}-e^{4\alpha}
	\frac{\partial V_Q}{\partial s_i}\right),
	\end{align}
	which explicitly shows the back-reaction of the quantum variables 
	$(s_i,p_{s_i},U_i)$ on the classical evolution.
	
	For later convenience, let us note that the full time-dependent effective potential term
	that appears in the Hamiltonian above can be rewritten in the following way:
	\begin{align}\label{quantum_potential_E}
	2e^{4\alpha}V_Q=
	&
	\frac{1}{3}
	\left(e^{4\alpha E_1}
	+e^{4\alpha E_2}+e^{4\alpha E_3}\right)
	-\frac{2}{3}
	\left(
	e^{4\alpha E_4}
	+e^{4\alpha E_5}+e^{4\alpha E_6}
	\right),
	\end{align}
	where the exponents $E_j$ ($j=1,2,3,4,5,6$) are defined as
	\begin{align}\label{exponents}
	\begin{split}
	&E_1=1-\frac{2\beta_+-8s_1^2}{\alpha},
	\\&
	E_2=1+\frac{\beta_++\sqrt{3}\beta_-+2s_1^2+6s_2^2}
	{\alpha},
	\\
	&
	E_3=1+\frac{\beta_+-\sqrt{3}\beta_-+2s_1^2+6s_2^2}
	{\alpha},
% 	\\
\end{split}
\begin{split}
	&
	E_4=1+\frac{\beta_++2s_1^2}{\alpha},
	\\
	&
	E_5=1+\frac{\sqrt{3}\beta_--\beta_++s_1^2+3s_2^2}
	{2\alpha},
	\\
	&
	E_6=1-\frac{\sqrt{3}\beta_-+\beta_+-s_1^2-3s_2^2}
	{2\alpha}.
	\end{split}
	\end{align}
	
	\section{A qualitative description of the dynamics}\label{sec:analysis_quantum}
	
	The dynamics of the classical Mixmaster model is characterized by a succession of Kasner regimes and transitions as the singularity is approached \cite{Mixmaster_I,Mixmaster}. 
	These regimes are defined as being kinetically dominated, in the sense that the potential term
	can be neglected in the Hamiltonian. Therefore, according to \eqref{hamiltonian}, the classical Hamiltonian
	is approximately given by the expression $H\approx(p_+^2+p_-^2)^{1/2}$, which corresponds to that of the exact Kasner model, and describes a free motion: the momenta $p_{+}$ and $p_-$ remain constant and the shape parameters
	$\beta_+$ and $\beta_-$ evolve as linear functions in $\alpha$. In the two-dimensional plane of the
	shape parameters, the trajectory is thus a straight line.
	During this evolution, however, at least one of the exponential terms of the potential
	\eqref{potential_Bianchi_IX} is growing in time and, from a certain
        point on, the potential is no longer negligible.
	The Kasner approximation then ceases to be valid, the system quickly bounces against the exponential walls
	depicted in Fig.~\ref{fig:equipotential}, and enters a new Kasner regime.
	The transition between successive Kasner epochs is described by a well-known
	map that relates the pre-bounce to post-bounce values of the momenta 
	$(p_{+}, p_-)$ and of the constants that characterize the Kasner evolution
	of the shape parameters $(\beta_+,\beta_-)$ (see, e.g., Ref. \cite{Belinski_book}).
	In fact, as mentioned in the introduction, this discrete map was already used
	to conclude that the Mixmaster model is chaotic in the early literature
	about this subject \cite{BKL,Barrow_2,Barrow_3,Barrow_4}.
	
	The described behavior is realized for a  generic choice of
          initial values.
	Nevertheless, there are certain specific values of the momenta during a Kasner regime, for which
	none of the exponential terms of the potential grows in time. The system thus never reaches an
	exponential wall and follows a Kasner evolution until it reaches the singularity. More precisely,
	there are three such possible values:
	$\{p_-<0,p_+=-\frac{p_-}{\sqrt{3}}\}$, $\{p_->0,p_+=\frac{p_-}{\sqrt{3}}\}$, and $\{p_-=0,p_+\leq 0\}$,
	which respectively correspond to the system going to infinity in the plane of the shape parameters 
	through one of the three vertices of the triangular shape of Fig. \ref{fig:equipotential}.
	These are sometimes called exits of the system. Given a set of initial data,
	some of them end up reaching one of these exits, while others follow
	an infinite succession of Kasner regimes and form the repeller set.
	This repeller has a fractal structure, which can be used to analyze
	the chaos of the system, as will be explained below.
	
	Let us now analyze how the described picture is modified by quantum effects.
	Numerical integration of the semiclassical evolution equations \eqref{equations_motion_quantum_beta}--\eqref{equations_motion_quantum_ps},
	not too close to the singularity and for relatively small values of the fluctuations,
	shows an evolution qualitatively similar to the classical one: the system spends
	most of the time following Kasner regimes, which are interrupted by rapid transitions
	when it hits the exponential walls.
	(For a recent analytic study on quantum Kasner transitions with a similar semiclassical
	model, we refer the reader to Ref.~\cite{Quantum_Bianchi_IX}).
	This can also be seen analytically. More precisely, 
	neglecting the potential (as well as its derivatives) in the equations of motion \eqref{equations_motion_quantum_beta}--\eqref{equations_motion_quantum_ps}, one can easily solve the system and obtain the following explicit evolution of the different variables during a Kasner epoch,
	\begin{align}\label{Kasner_quantum}
	\begin{split}
	&p_\pm=const.,
	\\
	&
	\beta_\pm=\frac{p_\pm}{P_Q}\alpha+c_\pm,
	\\
	&s_i=\sqrt{
		\frac{U_i}{B_i^2}
		+\frac{B_i^2}{P_Q^2}
		\left(\alpha+A_i\right)^2
	},
	\\
	&p_{s_i}=
	\frac{B_i^2\left(\alpha+A_i\right)}{\sqrt{
			U_iP_Q^2/B_i^2
			+B_i^2
			\left(\alpha+A_i\right)^2}},
	\end{split}
	\end{align}
	where $c_{\pm}$, $A_i$, and $B_i$ ($i=1,2$) are six integration
	constants. (See also Ref.~\cite{Description} for a discussion of free
	quantum dynamics in this language.)
	In particular $B_i$ encode the constant value of the fluctuation of the momenta:
	$\Delta(p_{+}^2)=p_{s_1}^2+U_1/s_1^2=B_1^2$, and $\Delta(p_{-}^2)=p_{s_2}^2+U_2/s_2^2=B_2^2$.
	The value of the Hamiltonian during these Kasner
	regimes is given by $P_Q:=\left(p_+^2+p_-^2+B_1^2+B_2^2\right)^{1/2}$; thus, quantum fluctuations
	increase its value with respect to its classical counterpart $P:=\left(p_+^2+p_-^2\right)^{1/2}$.
	The variables $p_{\pm}$ and $\beta_{\pm}$ evolve as in the classical model,
	the only difference being that the velocity of the anisotropies $\dot{\beta}_{\pm}$
	is a bit slower, since it inversely depends on $P_Q$. Towards the singularity,
	the variables $s_i$, which encode the fluctuations of the shape parameters, increase,
	while their momenta $p_{s_i}$ decrease, tending to their asymptotic values $-|B_i|$. 
	
	At this point, one can wonder about the fate of the classical exits mentioned above,
	where the system follows an infinite Kasner regime all the way to the singularity.
	It is clear from \eqref{quantum_hamiltonian_approx} that,
	for such an exit to exist for the quantum system,
	all the exponents $\{E_j\}_{j=1}^6$ in \eqref{exponents} must be nonnegative
	during the whole Kasner evolution given by \eqref{Kasner_quantum}.
	As a representative example, let us analyze the behavior of the first exponent, $E_1$.
	During a Kasner regime, this exponent evolves as
	\begin{align}\label{exp_1_kasner}
	E_1=\frac{P_Q-2p_+}{P_Q}-\frac{2c_+}{\alpha}+\frac{8U_1}{B_1^2\alpha}+\frac{8B_1^2}{P_Q^2}\frac{(\alpha+A_1)^2}{\alpha}.
	\end{align}
	The first two terms are the only ones that appear in the classical limit: $(P_Q-2p_+)/P_Q$ is constant, and $-2c_+/\alpha$ vanishes as $\alpha\to -\infty$. Therefore, under the classical evolution, the sign of $(P_Q-2p_+)/P_Q$
	rules the asymptotic behavior of this exponential term. In particular, if it is negative,
	it will define the usual ``classical'' potential wall asymptotically, with an exponential shape $e^{-m^2\alpha}$, which will eventually
	put an end to the Kasner dynamics. However, if it is vanishing, the system will be following
	one of the exit trajectories (under the additional condition that all the other exponents $E_j$ are nonnegative). Finally, if it is positive, $E_1$ will provide a negligible contribution, and
	another exponential term $E_j$ will dominate and define the exponential wall as $\alpha\to -\infty$.
	
	Quantum contributions introduce further $\alpha$-dependence on the exponent $E_1$.
	In particular, the term with $U_1$ tends to zero as the singularity is approached,
	and thus it does not change the classical asymptotic picture. However, the last term,
	$8B_1^2P_Q^{-2}(\alpha+A_1)^2/\alpha$, increases as
	$\alpha\to-\infty$ (just like the variance of a free-particle wave
        function). Since it is determined by quantum fluctuations, this
          term is initially small in a semiclassical regime. But if the system
          is on (or close to) a classical exit trajectory and does not bounce
          before this term becomes relevant, it will eventually dominate in
        the expression for $E_1$.
	In such a case, this term will effectively define a potential of the form $e^{k^2\alpha^2}$,
	breaking the Kasner evolution.
	
	In fact, performing the same analysis for the different exponents $E_j$, there is always
	a contribution from a term proportional to the relative fluctuation
        $B_i^2/P_Q^2$ that implies a negative value for
	the corresponding exponent $E_j$ as $\alpha\to-\infty$. More specifically,
	the Kasner dynamics of the different exponents read as follows,
	\begin{align}\label{exponents_kasner}
	\begin{split}
	&E_2=\frac{P_Q+p_++\sqrt{3}p_-}{P_Q}+
	\frac{c_++\sqrt{3}c_-}{\alpha}
	+\frac{2U_1}{B_1^2\alpha}
	+\frac{6U_2}{B_2^2\alpha}
	+\frac{2B_1^2}{P_Q^2\alpha}(\alpha+A_1)^2
	+\frac{6B_2^2}{P_Q^2\alpha}(\alpha+A_2)^2,
	\\
	&E_3=\frac{P_Q+p_+-\sqrt{3}p_-}{P_Q}+
	\frac{c_+-\sqrt{3}c_-}{\alpha}
	+\frac{2U_1}{B_1^2\alpha}
	+\frac{6U_2}{B_2^2\alpha}
	+\frac{2B_1^2}{P_Q^2\alpha}(\alpha+A_1)^2
	+\frac{6B_2^2}{P_Q^2\alpha}(\alpha+A_2)^2,
	\\
	&E_4=\frac{P_Q+p_+}{P_Q}+
	\frac{c_+}{\alpha}
	+\frac{2U_1}{B_1^2\alpha}
	+\frac{2B_1^2}{P_Q^2\alpha}(\alpha+A_1)^2,
	\\
	&E_5=\frac{2P_Q-p_++\sqrt{3}p_-}{2P_Q}+
	\frac{\sqrt{3}c_--c_+}{2\alpha}
	+\frac{U_1}{2B_1^2\alpha}
	+\frac{3U_2}{2B_2^2\alpha}
	+\frac{B_1^2}{2P_Q^2\alpha}(\alpha+A_1)^2
	+\frac{3B_2^2}{2P_Q^2\alpha}(\alpha+A_2)^2,
	\\
	&E_6=\frac{2P_Q-p_+-\sqrt{3}p_-}{2P_Q}-
	\frac{\sqrt{3}c_-+c_+}{2\alpha}
	+\frac{U_1}{2B_1^2\alpha}
	+\frac{3U_2}{2B_2^2\alpha}
	+\frac{B_1^2}{2P_Q^2\alpha}(\alpha+A_1)^2
	+\frac{3B_2^2}{2P_Q^2\alpha}(\alpha+A_2)^2.
	\end{split}
	\end{align}
	Therefore, and contrary to the classical case, there is no combination
	of parameters that would lead to all $E_j$ being simultaneously nonnegative as $\alpha\to-\infty$.
	All the classical exits are thus closed by quantum fluctuations.
	
	We will now estimate the ranges of Kasner parameters that characterize
the conditions for the system to hit one of those quantum walls.
	Let us assume a trajectory that is following a Kasner dynamics
	with small fluctuations relatively close to the singularity. That is,
	$\alpha$ is already large enough (in absolute value) so that all the terms of the form
	$1/\alpha$ can be neglected in \eqref{exponents_kasner},
	but not so large yet as to make terms proportional to $B_i^2/P_Q^2$ relevant.
	In such a scenario, all the dependence on $\alpha$ can be neglected in
	the exponentials \eqref{exponents_kasner}. If one then requires all of them to be nonnegative,
	so that none of the potential terms is growing along evolution,
	the following three ranges of values are obtained,
	\begin{align}\label{exits}
	\begin{split}
	p_-\geq 0\quad\text{and}\quad
	\frac{1}{4}\left(\sqrt{9-3r}-\sqrt{1+r}\right)
	\leq \frac{p_+}{(p_+^2+p_-^2)^{1/2}}\leq
	\frac{1}{2}\sqrt{1+r}
	, 
	\\
	p_-<0\quad\text{and}\quad
	\frac{1}{4}\left(\sqrt{9-3r}-\sqrt{1+r}\right)
	\leq \frac{p_+}{(p_+^2+p_-^2)^{1/2}}\leq
	\frac{1}{2}\sqrt{1+r},
	\\
	\quad
	\frac{p_+}{(p_+^2+p_-^2)^{1/2}} \leq
	-\frac{1}{4}\left(\sqrt{9-3r}+\sqrt{1+r}\right),
	\end{split}
	\end{align}
	where $r:=(B_1^2+B_2^2)/(p_+^2+p_-^2)$.  If the parameters of the
        system lie in any of these ranges, the system will follow its Kasner
        dynamics without hitting any classical potential wall until the
        fluctuations $B_i^2/P_Q^2$ become dominant. This is when the system
        bounces against a quantum wall.  Note that in the classical limit
        $r\to 0$, the above ranges correspond to the classical exits,
        $\{p_-<0,p_+=-\frac{p_-}{\sqrt{3}}\}$,
        $\{p_->0,p_+=\frac{p_-}{\sqrt{3}}\}$, and $\{p_-=0,p_+\leq 0\}$,
        respectively.
	
	Intuitively the above relations define three wedges on the plane of anisotropies.
	The classical convex potential walls depicted in Fig. \ref{fig:equipotential}
	are pushed back from the three symmetry axes by the quantum effects
	creating such wedges. These wedges do not reach infinity, though,
	as they are closed by the new quantum potential walls produced by the fluctuations $B_i^2/P_Q^2$.
	Therefore, these quantum contributions form concave caps
	connecting classical walls, and constraining the dynamics of the system to a finite region of
	the phase space.
	
	\section{Analysis of the chaos}
	\label{sec:analysis_chaos}
	
	As explained in the introduction, the classical Mixmaster model shows
        a chaotic behavior close to the singularity.  In the context of
        general relativity, and specifically for the Mixmaster model, two
        methods have been proposed in the literature to study the chaotic
        nature of this system without ambiguities.  The first one makes use of
        the covariantly defined Lyapunov exponent
        \cite{Montani_paper,Montani_book}, while the second one is based on
        the study of the fractal structure of the space of initial conditions
        \cite{Cornish_Levin}.  In the following, both methods will be applied
        to our semiclassical system in order to check how quantum fluctuations
        modify the classical results. More precisely, in
        Sec.~\ref{sec:Lyapunov} we will present a canonical transformation to
        a set of variables that generalize the Misner-Chitre variables, which
        will allow us to compute the Lyapunov exponent, and construct an
        isomorphism between the quantum Mixmaster dynamics and the geodesic
        flow on a four-dimensional curved manifold.  The conclusion of this
        method will be that the quantum dynamics is still chaotic.  In
        Sec.~\ref{sec:fractal} we will be able to provide a quantitative
         measure of the
        chaos by making use of the fractal method, and will show that, as
        compared to the classical system, the quantum system has a reduced
        degree of  chaos.
	
	\subsection{Covariant Lyapunov exponent}
	\label{sec:Lyapunov}
	
	The Lyapunov exponent is an accurate way of representing chaos in dynamical
	systems
	as long as the following conditions are satisfied \cite{Motter_lyap}:
	\begin{enumerate}
		\item The system is autonomous.
		\item The relevant part of the phase space is bounded.
		\item The invariant measure is normalizable.
		\item The domain of the time parameter is infinite.
	\end{enumerate}
	
	The classical dynamics of the Mixmaster model described by the shape parameters
	$\beta_{\pm}$ and their conjugate momenta $p_\pm$ in terms of the internal time $\alpha$
	does not obey all of the above conditions.
	Namely, while the domain of the time parameter $\alpha$ is infinite,
	the system \eqref{equations_motion_classical_beta_+}--\eqref{equations_motion_classical_p_-} is not autonomous due to the explicit term $e^{4\alpha}$.
	Moreover, due to the appearance of this same factor in the Hamiltonian \eqref{hamiltonian},
	asymptotically the height of the potential walls is effectively diminished
	making the accessible region of the phase space unbounded.
	However, for the classical model one can perform a canonical
	transformation to a new set of variables introduced by Misner and Chitre~\cite{Montani_paper, Chitre},
	which defines an asymptotically time-independent potential term in the Hamiltonian.
	In terms of these variables the potential walls are stationary, defining a bounded
	region of the phase space for the dynamics of the system, and can be approximated
	by infinite potential walls fixed at certain positions. Furthermore, the dynamics
	is shown to be isomorphic to the geodesic flow of a curved two-dimensional Riemannian manifold.
	This picture is very helpful and allows, in particular, to compute the corresponding
	Lyapunov exponent by analyzing the geodesic deviation equation on that manifold.
	
	Here, we will generalize the classical analysis by constructing a canonical
	transformation for our quantum system \eqref{equations_motion_quantum_beta}--\eqref{equations_motion_quantum_ps}, which will lead to a new set of variables
	that obey the properties detailed above. Furthermore, an isomorphism
	to the geodesic motion on a four-dimensional Riemannian manifold will be obtained, and the
	corresponding Lyapunov exponent will be computed.
	
	\subsubsection{Generalized Misner-Chitre variables}
	
	Let us start by implicitly defining the generalized Misner-Chitre variables $(\Gamma,\xi,\theta,\sigma,\phi)$ by the following transformation,
	\begin{align}\label{misner-chitre}
	\begin{split}
	\alpha&=-e^{\Gamma}\xi,
	\\
	\beta_+&=e^{\Gamma}\sqrt{\xi^2-1}\cos\theta,
	\\
	\beta_-&=e^{\Gamma}\sqrt{\xi^2-1}\sin\theta\cos\sigma,
	\\
	s_1&=e^{\Gamma}\sqrt{\xi^2-1}\sin\theta\sin\sigma f_+(\phi,p_\phi),
	\\
	s_2&=e^{\Gamma}\sqrt{\xi^2-1}\sin\theta\sin\sigma f_-(\phi, p_\phi),
	\end{split}
	\end{align}
	for configuration variables, and their conjugate momenta $(p_\Gamma,p_\xi,p_\theta,p_\sigma,p_\phi)$ by the relations
	\begin{align}
	\nonumber
	p_{\alpha}&=e^{-\Gamma}\left[
	(\xi^2-1)p_{\xi}-\xi p_{\Gamma}
	\right],
	\\
	\nonumber
	p_{+}&=e^{-\Gamma}\left[
	\cos\theta\sqrt{\xi^2-1}(
	\xi p_\xi-p_{\Gamma}
	)
	-\frac{\sin\theta}{\sqrt{\xi^2-1}}p_\theta
	\right],
	\\\label{misner-chitre-moments}
	p_-&=e^{-\Gamma}\left[
	\sin\theta\cos\sigma\sqrt{\xi^2-1}(
	\xi p_\xi-p_{\Gamma}
	)
	+\frac{\cos\theta\cos\sigma}{\sqrt{\xi^2-1}}p_\theta
	-\frac{\csc\theta\sin\sigma}{\sqrt{\xi^2-1}}p_\sigma
	\right],
	\\
	\nonumber
	p_{s_1}&=e^{-\Gamma}f_+(\phi,p_\phi)
	\left[\sin\sigma
	\sin\theta \sqrt{\xi^2-1}
	(
	\xi p_\xi-p_{\Gamma}
	)
	+\frac{\cos\theta }{\sqrt{\xi^2-1}}p_\theta
	+\frac{\csc\theta\cos\sigma}{\sqrt{\xi^2-1}}p_\sigma
	-\frac{h(p_\phi) \csc\theta\csc\sigma\cos{2\phi}}{2p_\phi f_+^2(\phi,p_\phi)\sqrt{\xi^2-1}}
	\right],
	\\
	p_{s_2}&=e^{-\Gamma}f_-(\phi,p_\phi)
	\left[
	\sin\theta\sin\sigma \sqrt{\xi^2-1}
	(
	\xi p_\xi-p_{\Gamma}
	)
	+\frac{\cos\theta\sin\sigma}{\sqrt{\xi^2-1}}p_\theta
	+\frac{\csc\theta\cos\sigma}{\sqrt{\xi^2-1}}p_\sigma
	+\frac{h(p_\phi)\csc\theta\csc\sigma\cos{2\phi}}{2p_\phi f_-^2(\phi,p_\phi)\sqrt{\xi^2-1}}
	\right],
	\nonumber
	\end{align}
where the functions $f_+(\phi,p_{\phi})$, $f_-(\phi,p_\phi)$, and $h(p_\phi)$ are defined as follows,
	\begin{align}
	\label{f_+_phi}
	&f_+(\phi,p_\phi):=\frac{1}{\sqrt{2}|p_\phi|}\left(
	p_\phi^2+U_1-U_2-h(p_\phi)\sin{2\phi}
	\right)^{1/2},
	\\	\label{f_-_phi}
	&f_-(\phi, p_\phi):=\frac{1}{\sqrt{2}|p_\phi|}\left(
	p_\phi^2-U_1+U_2+
	h(p_\phi)\sin{2\phi}
	\right)^{1/2},
	\\	\label{h_phi}
	&h(p_\phi):=\sqrt{
		\left(p_\phi^2-(U_1+U_2)\right)^2-4U_1U_2}.
	\end{align}
		The domain of definition of the different variables is given by $\theta,\sigma\in(0,\pi)$,
		$\xi\geq 1$, $p_\xi,p_\theta,p_\sigma,p_\Gamma,\Gamma\in\mathbb{R}$, while the pair $(\phi, p_\phi)$
takes values in the ranges $\phi\in[0,2\pi)$, and $p_\phi\in\mathbb{R}$ is restricted under the condition
that the above functions $f_+$, $f_-$, and $h$ are real.
	For the classical system, only the variables $(\Gamma,\xi, \theta)$ are
defined, along with their corresponding conjugate momenta
$(p_\Gamma,p_\xi, p_\theta)$. In order to encode the quantum degrees of
freedom, the new angular variables $(\sigma, \phi)$ and their momenta
$(p_\sigma, p_\phi)$ have been introduced. The classical configuration space corresponds to the union of the two half-planes $\sigma\to 0$ and $\sigma\to \pi$. Note that points in this plane are unreachable for the quantum system since $s_1$ and $s_2$ would be vanishing there, violating the uncertainty principle.
This does not only apply to the classical plane, but also to other sets of points such as those given by $\sin\theta=0$, $f_+(\phi,p_\phi)=0$, or $f_-(\phi,p_\phi)=0$.

This canonical
transformation is parameterized by the constants of motion $U_1$ and $U_2$,
which encode information about the saturation of the uncertainty
relation. If $U_1$ and $U_2$ could be ignored, the functions \eqref{f_+_phi}--\eqref{h_phi}
would simplify to $f_+= |\cos(\phi+\pi/4)|$,
  $f_-= |\sin(\phi+\pi/4)|$, and $h=p_{\phi}^2$. The configuration space mapping
  (\ref{misner-chitre}) would then be a direct generalization of 2-dimensional polar
  coordinates on the classical anisotropy plane to a 4-dimensional quantum
  version. Inclusion of the $U$-terms requires a deformation of the 3-sphere
  mapping according to \eqref{f_+_phi}--\eqref{f_-_phi}. The functions $f_+$ and $f_-$ are then
  nonzero because $U_1$ and $U_2$ are positive, and so are $s_1$ and $s_2$
  due to the restrictions of $\theta$ and $\sigma$ to the range
  $(0,\pi)$. Note also that the condition that $h(p_{\phi})$ be real
  imposes a lower bound on $|p_{\phi}|$ for nonzero $U_1$ and $U_2$. The
  original repulsive terms $U_1/s_1^2$ and $U_2/s_2^2$ in the effective
  potential are now replaced by a lower bound on the angular momentum in the
  $s_1$-$s_2$ plane.

If we now choose $\Gamma$ as the internal time variable, the Hamiltonian that governs the dynamics of the
remaining variables $(\xi,\theta,\sigma,\phi)$ with respect to $\Gamma$ is given by
\begin{align}\label{quantum_hamiltonian_misner_chitre}
\mathcal{H}=\left[
p_{\xi}^2(\xi^2-1)+\frac{p_{\theta}^2}{\xi^2-1}+
\frac{p_{\sigma}^2}{(\xi^2-1){\sin^2\theta}}
+
\frac{p_{\phi}^2}{(\xi^2-1){\sin^2\theta}\sin^2\sigma}
+2e^{2\Gamma-4\xi e^{\Gamma}}V_Q
\right]^{1/2}.
\end{align}
The potential term is obtained by simply applying the above canonical
transformation to its definition \eqref{quantum_potential},
and its full contribution to the Hamiltonian can be written as
a linear combination of exponential terms,
\begin{equation}\label{potential_misner_chitre}
2e^{2\Gamma-4\xi e^{\Gamma}}V_Q=
	\frac{e^{2\Gamma}}{3}
	\left(e^{-4\xi e^{\Gamma} E_1}
	+e^{-4\xi e^{\Gamma} E_2}+e^{-4\xi e^{\Gamma} E_3}\right)
	-\frac{2}{3}e^{2\Gamma}
	\left(
	e^{-4\xi e^{\Gamma} E_4}
	+e^{-4\xi e^{\Gamma} E_5}+e^{-4\xi e^{\Gamma} E_6}
	\right),
\end{equation}
with the same exponents $\{E_i\}_{i=1}^6$ defined above \eqref{exponents}.
In terms of the new variables, these exponents take the explicit form
\begin{align}\label{exponents_Misner_Chitre}
	\begin{split}
	&E_1=1+\frac{2\sqrt{\xi^2-1}}{\xi}\cos\theta
	-\frac{8e^{\Gamma}(\xi^2-1)}{\xi}\sin^2\theta\sin^2\sigma f_+^2(\phi,p_\phi),
	\\
	&E_2=1-\frac{\sqrt{\xi^2-1}}{\xi}(\cos\theta+\sqrt{3}\sin\theta\cos\sigma)
	-\frac{2e^{\Gamma}(\xi^2-1)}{\xi}\sin^2\theta\sin^2\sigma[f_+^2(\phi,p_\phi)+3f_-^2(\phi,p_\phi)],
		\\
	&E_3=1-\frac{\sqrt{\xi^2-1}}{\xi}(\cos\theta-\sqrt{3}\sin\theta\cos\sigma)
	-\frac{2e^{\Gamma}(\xi^2-1)}{\xi}\sin^2\theta\sin^2\sigma[f_+^2(\phi,p_\phi)+3f_-^2(\phi,p_\phi)],
	\\
	&E_4=1-\frac{\sqrt{\xi^2-1}}{\xi}\cos\theta
	-\frac{2e^{\Gamma}(\xi^2-1)}{\xi}\sin^2\theta\sin^2\sigma f_+^2(\phi,p_\phi),
	\\
	&E_5=1+\frac{\sqrt{\xi^2-1}}{2\xi}(\cos\theta-\sqrt{3}\sin\theta\cos\sigma)
	-\frac{e^{\Gamma}(\xi^2-1)}{2\xi}\sin^2\theta\sin^2\sigma[f_+^2(\phi,p_\phi)+3f_-^2(\phi,p_\phi)],
		\\
	&E_6=1+\frac{\sqrt{\xi^2-1}}{2\xi}(\cos\theta+\sqrt{3}\sin\theta\cos\sigma)
	-\frac{e^{\Gamma}(\xi^2-1)}{2\xi}\sin^2\theta\sin^2\sigma[f_+^2(\phi,p_\phi)+3f_-^2(\phi,p_\phi)].
	\end{split}
\end{align}
The singularity is now located at $\Gamma\rightarrow\infty$,
which ensures the fulfillment of condition 4 above,
and thus
the potential \eqref{potential_misner_chitre} will be asymptotically negligible
as long as all exponential terms \eqref{exponents_Misner_Chitre} are positive.
In the classical case, which can be recovered simply by imposing $\sin\sigma\to0$
in \eqref{exponents_Misner_Chitre}, all these exponents turn out to be independent
of $\Gamma$. Therefore, $\Pi$, defined as the region of the configuration space
where the potential is negligible, can be shown to be compact and time independent.
Outside this region, the potential will tend to infinity and the system will
be prevented from being located there. In this way,
the potential can be asymptotically approximated by certain combination
of stationary walls and condition 2 above is met.

Quantum contributions, however, render the above exponents $E_i$ time dependent.
Therefore, the shape of the region $\Pi$ will also change with $\Gamma$.
The key issue to check is whether such a dependence could
make the region $\Pi$ noncompact and spoil the good properties of the classical
Misner-Chitre variables. The answer, though, is in the negative.
As can be easily checked in the explicit expression of the exponents
\eqref{exponents_Misner_Chitre}, all quantum terms have a negative contribution,
so their relevance increases as the singularity is approached.
Thus, as $\Gamma$ tends to infinity, quantum contributions
decrease the value of every exponent $E_i$ and, in this way,
make the region $\Pi$ shrink. In fact,
except in the planes defined by the values $ \sin\sigma\to0$ and  $\sin\theta\to0$,
from certain value of $\Gamma$ on, the relevance of the quantum effects
will be such that all the exponents $E_i$ will be negative. Therefore, outside the mentioned planes, the potential term
will not be negligible anywhere and the region $\Pi$ will be restricted to the sets of points defined by $\sin\theta\to 0$ and $\sin\sigma\to 0$, which are forbidden for the quantum system.
This is related to the closed-off classical exits mentioned above: while
in the classical system there are certain values of the parameters
which avoid the potential all along until the singularity, in
the quantum system any value of the parameters will lead, sooner or later, to an interaction
with the potential.

In the strict limit of $\Gamma\to\infty$ the region $\Pi$ is empty, but
  any finite range of large values of $\Gamma$ implies a nonempty and bounded
  set. In any such region, independently of the specific finite range of
  $\Gamma$, the Hamiltonian in terms of the generalized Misner-Chitre variables
is given by just the kinetic terms, namely,
\begin{align}\label{Hamiltonian_pi}
&	\mathcal{H}=\left[p_{\xi}^2(\xi^2-1)+\frac{p_{\theta}^2}{\xi^2-1}+
\frac{p_{\sigma}^2}{(\xi^2-1){\sin^2\theta}}
+
\frac{p_{\phi}^2}{(\xi^2-1){\sin^2\theta}\sin^2\sigma}
\right]^{1/2}.
\end{align}
Since it does not depend explicitly on time, it is constant under evolution
and defines the conserved energy of the dynamical system $E={\cal H}$.
Consequently, the system of equations of motion
\eqref{equations_motion_quantum_beta}--\eqref{equations_motion_quantum_ps}
written for the new variables becomes autonomous, and condition 1 above is
fulfilled.

As we already showed, by introducing our new set of variables, we have been
able to restrict the relevant part of the phase space to a bounded region
$\Pi$, and thus condition 2 of the above criteria is satisfied. Moreover, in
this same region and asymptotically, following the same rationale as in the
classical case \cite{Montani_book}, one can conclude that the invariant
measure is normalizable, satisfying thus also condition 3. Finally,
following condition 4, the domain of the time parameter $\Gamma$ is infinite,
% the local dynamics is the same in any region $\Pi$ for a given finite range of
%   $\Gamma$, and the overall domain of $\Gamma$ is infinite,
  all conditions 1--4 listed above are met. The sign of the Lyapunov exponent that can be
computed for these variables is thus invariant, and will characterize the
possible chaos of the system.

\subsubsection{Isomorphism to a geodesic flow on a Riemannian manifold}

In order to compute the Lyapunov exponent, it is very useful to note that the
above Hamiltonian \eqref{Hamiltonian_pi} has a form similar to the Hamiltonian
$(\frac{1}{2}g_{\mu\nu}p^\mu p^\nu)^{1/2}$ of a free particle evolving on a
curved background with metric $g_{\mu\nu}$.  Therefore, asymptotically (for
  any finite range of large values of $\Gamma$, as explained in the preceding
  subsection), the Mixmaster dynamics inside the exponential walls is
isomorphic to the geodesic flow on the 4-dimensional Riemannian manifold
with metric
\begin{align}\label{metric_riemann}
g_{\mu\nu} dx^\mu dx^\nu=E^2\left[
\frac{d\xi ^2}{\xi^2-1}+\left(\xi^2-1\right)
d\theta^2
+\left(\xi^2-1\right)\sin\theta^2
d\sigma^2
+\left(\xi^2-1\right)\sin\theta^2\sin\sigma^2
d\phi^2
\right].
\end{align}
[The coordinate transformation $\xi=\cosh\psi$ shows that this space is
  a submanifold of a hyperboloid $|\vec{x}|^2-t^2=E^2$ in 5-dimensional Minkowski space-time with line element $ds^2= E^2(d\psi^2+ \sinh^2\psi(d\theta^2+\sin^2\theta
  (d\sigma^2+\sin^2\sigma d\phi^2)))$.]
This metric is maximally symmetric, with curvature scalar $R=-12/E^2$, and
thus its Riemann tensor can be written as
\begin{align}\label{riemann}
{R_{\tau\rho\sigma}}^ {\mu}=
\frac{R}{12}\left(
g_{\tau\sigma}g^{\mu}_{\rho}-
g_{\rho\sigma}g^{\mu}_{\tau}
\right)=
-\frac{1}{E^2}
\left(
g_{\tau\sigma}g^{\mu}_{\rho}-
g_{\rho\sigma}g^{\mu}_{\tau}
\right).
\end{align}
Moreover, from this expression, it is easy to compute the sectional curvature of the manifold.
That is, given any two linearly independent vector fields $a^{\mu}$ and $b^{\mu}$, we have
\begin{align}\label{sectional_curvature}
\frac{{R_{\tau\rho\sigma\nu}}
	a^{\tau}b^{\rho}a^{\sigma}b^{\nu}}
{(a^{\mu}a_{\mu})(b^{\alpha}b_{\alpha})-(a^{\mu}b_{\mu})^2}
=
-\frac{1}{E^2}.
\end{align}
Therefore, the sectional curvature is constant and negative.

In order to obtain the invariant Lyapunov exponent of the system, one needs to analyze the geodesic
deviation on this manifold. For such a purpose, it is very convenient to introduce the Fermi orthonormal basis,
\begin{align}\label{fermi_basis}
\begin{split}
&e^{\mu}_1=\left(
\frac{\sqrt{\xi^2-1}}{E}\cos\gamma_{1}(s),
\frac{\sin\gamma_{1}(s)\cos\gamma_{2}(s)}{E\sqrt{\xi^2-1}},
\frac{\sin\gamma_{1}(s)\sin\gamma_{2}(s)\cos\gamma_{3}(s)}{E\sqrt{\xi^2-1}\sin\theta},
\frac{\sin\gamma_{1}(s)\sin\gamma_{2}(s)\sin\gamma_{3}(s)}{E\sqrt{\xi^2-1}\sin\theta\sin\sigma}
\right)
,
\\
&e^{\mu}_2=\left(
-\frac{\sqrt{\xi^2-1}}{E}\sin\gamma_{1}(s),
\frac{\cos\gamma_{1}(s)\cos\gamma_{2}(s)}{E\sqrt{\xi^2-1}},
\frac{\cos\gamma_{1}(s)\sin\gamma_{2}(s)\cos\gamma_{3}(s)}{E\sqrt{\xi^2-1}\sin\theta},
\frac{\cos\gamma_{1}(s)\sin\gamma_{2}(s)\sin\gamma_{3}(s)}{E\sqrt{\xi^2-1}\sin\theta\sin\sigma}
\right),
\\
&e^{\mu}_3=\left(
0,
-\frac{\sin\gamma_{2}(s)}{E\sqrt{\xi^2-1}},
\frac{\cos\gamma_{2}(s)\cos\gamma_{3}(s)}{E\sqrt{\xi^2-1}\sin\theta},
\frac{\cos\gamma_{2}(s)\sin\gamma_{3}(s)}{E\sqrt{\xi^2-1}\sin\theta\sin\sigma}
\right),
\\
&e^{\mu}_4=\left(
0,0,
-\frac{\sin\gamma_{3}(s)}{E\sqrt{\xi^2-1}\sin\theta},
\frac{\cos\gamma_{3}(s)}{E\sqrt{\xi^2-1}\sin\theta\sin\sigma}
\right),
\end{split}
\end{align}
so that $g_{\mu\nu}e_i^\mu e_j^{\nu}=\delta_{ij}$,
where $\gamma_1,\gamma_2,\gamma_3\in[0,2\pi)$ are angular variables that depend on the curvilinear coordinate $s$. These variables are then fixed by requiring the first vector $e^{\mu}_1=\left(
d\xi/ds,
d\theta/ds,
d\sigma/ds,
d\phi/ds
\right)$ to be tangent to the geodesics,
\begin{align}
\label{u_geodesic_2}
\frac{De^{\mu}_1}{ds}&:=
\frac{de^{\mu}_1}{ds}+\Gamma^{\mu}{}_{\nu\rho}e_1^{\nu}e_1^{\rho}=0,
\end{align}
where $\Gamma^{\mu}{}_{\nu\rho}$ are the Christoffel symbols of the metric \eqref{metric_riemann},
which leads to the equations
\begin{align}\label{cond_angles_geodesic}
\begin{split}
&\frac{d\gamma_1(s)}{ds}=
-\frac{\xi(s)\sin\gamma_1(s)}{E\sqrt{\xi(s)^2-1}},
\\
&\frac{d\gamma_2(s)}{ds}=
-\frac{\sin\gamma_1(s)\sin\gamma_2(s)}{E\sqrt{\xi(s)^2-1}\tan\theta(s)},
\\
&\frac{d\gamma_3(s)}{ds}=
-\frac{\sin\gamma_1(s)\sin\gamma_2(s)
	\sin\gamma_3(s)}{E\sqrt{\xi(s)^2-1}\tan\sigma(s)\sin\theta(s)}.
\end{split}
\end{align}
The basis above is constructed so that all the vectors are parallel transported
along this curve and, thus,
\begin{equation}
\frac{De^{\mu}_i}{ds}=
\frac{de^{\mu}_i}{ds}+\Gamma^{\mu}{}_{\nu\rho}
e_1^{\nu}e_i^{\rho}=0
\end{equation}
is obeyed for all $i=1,2,3,4$.

Let us now consider the vector $Z^{\mu}$ that connects two nearby geodesic curves,
and obeys the geodesic deviation equation,
\begin{align}\label{geodesic_deviation}
\frac{D^2Z^{\mu}}{ds^2}={R_{\tau\rho\sigma}}^ {\mu}e_1^{\sigma}e_1^{\rho}Z^{\tau},
\end{align} 
with ${R_{\tau\rho\sigma}}^{\mu}$ being the Riemann tensor \eqref{riemann}.
Decomposing this vector in the above basis,
\begin{equation}
Z^{\mu}=Z_i \delta^{ik} e_k^{\mu}, 
\end{equation}
with $Z_i\in\mathbb{R}$,
and taking into account the properties of the basis vectors $e^\mu_i$,
the geodesic deviation equation takes the form
\begin{align}
\label{geodesic_deviation_simply}
\frac{d^2Z_1}{ds^2}e^{\mu}_1+\frac{d^2Z_2}{ds^2}e^{\mu}_2+\frac{d^2Z_3}{ds^2}e^{\mu}_3+\frac{d^2Z_4}{ds^2}e^{\mu}_4
=\frac{1}{E^2}\left(Z_2 e^{\mu}_2+Z_3 e^{\mu}_3+Z_4 e^{\mu}_4
\right).
\end{align}
Since these basis vectors are linearly independent, their coefficients on both sides of the equation must be identical, that is,
\begin{align}
\label{deviation_equation}
\frac{d^2Z_1}{ds^2}=0,
\qquad    
\frac{d^2Z_2}{ds^2}=\frac{Z_2}{E^2},
\qquad
\frac{d^2Z_3}{ds^2}=\frac{Z_3}{E^2},
\qquad
\frac{d^2Z_4}{ds^2}=\frac{Z_4}{E^2}.
\end{align}
It is straightforward to obtain the solutions to these equations,
\begin{align}\label{solution_deviation}
Z_1=C_1s+D_1,\qquad
Z_2=C_2e^{s/E}+D_2e^{-s/E},
\qquad
Z_3=C_3e^{s/E}+D_3e^{-s/E},
\qquad
Z_4=C_4e^{s/E}+D_4e^{-s/E},
\end{align}
with integration constants $C_i$ and $D_i$ for $i=1,2,3,4$.
Then, according to \cite{Montani_paper}, the invariant Lyapunov exponent of this system is given as
\begin{align}\label{lyapunov_geo}
\lambda=\sup_{i=1,2,3,4}\left\{
\lim_{s\to\infty}\frac{\ln\left[
	Z_i^2+\left(dZ_i/ds\right)^2
	\right]}{2s}
\right\}.
\end{align}
Substituting the above solutions \eqref{solution_deviation} in this definition,
the invariant Lyapunov exponent is found to be
\begin{align}\label{lyapunov_final}
\lambda=\frac{1}{E},
\end{align}
which is strictly positive.

As already explained, since all the conditions 1-4 detailed above are satisfied by this system,
the positive value of this Lyapunov exponent \eqref{lyapunov_final} implies
that the dynamics
under consideration is chaotic. Since this geodesic flow is isomorphic to the quantum dynamics
of the Mixmaster model, we can also confirm that the latter is chaotic.

\subsection{The fractal dimension of the repeller}\label{sec:fractal}

If a dynamical system has a finite number of exits or possible final outcomes (for instance, in the presence of an attractor),
the space of initial conditions can be divided into different regions according to the final state of each point.
The boundary between these regions
can be either smooth or fractal, and the latter, as will be shown below, is a clear indicator of chaos.
Given the dimension of the boundary $D_0$, and
the dimension of the space of initial data $D$, we define its difference as
\begin{equation}\label{uncertainty_coeff_D}
\delta:=D-D_0.
\end{equation}
When the boundary is smooth, we have $\delta=1$, while for a fractal structure,
$\delta$ is in the range $0<\delta<1$. This quantity is known as the uncertainty exponent,
since it is directly related to the fraction of the space of initial data with uncertain outcome
\cite{Cornish_Levin,mcdonald}. Specifically,
\begin{equation}\label{uncertainty_coeff}
f(\varepsilon)\sim\varepsilon^{\delta},
\end{equation}
where $\varepsilon$ is the error of the initial conditions, and $f(\varepsilon)$ is
the fraction of uncertain outcomes in the space of initial data.
Here, a point is classified as being of uncertain outcome for an error $\varepsilon$
if, taking a hypersphere of radius $\varepsilon$ centered around that point,
it contains points with different outcomes.
Nonchaotic systems do not present an amplification of the initial uncertainties, and, correspondingly,
they show a linear relation between $f(\varepsilon)$ and $\varepsilon$.
Therefore, according to \eqref{uncertainty_coeff}, $\delta=1$, which, in turn, implies that the
boundary between the different regions is smooth. Chaotic systems, on the contrary, show
an amplification of the initial error, which means that $0<\delta<1$,
and that the boundaries are fractal.
The lower the value of $\delta$ (and thus higher the value of $D_0$), the more
chaotic the system is \cite{Cornish_Levin}.

Concerning the analysis of theories with time-reparameterization invariance,
like general relativity, the advantage of measuring the chaos in this way, as
opposed to other dynamical techniques, is that $\delta$ is invariant.  Indeed,
a reparameterization of the time variable will not change the
classification---in terms of different outcomes---of each point on the space
of initial conditions. An initial point that has a given outcome will
  maintain the very same outcome under a time reparameterization, and only
  reach it earlier or later in the new time coordinate.  Therefore,
since $\delta$ is a measurable quantity, it can be used to compare the level
of chaos between different systems, and this is certainly an appropriate
method to study how quantum effects modify the chaotic behavior of the
classical Mixmaster model.

\subsubsection{Application to the Mixmaster model and numerical setup}

The first step in order to apply this framework to our model, and compute
the uncertainty exponent $\delta$, is thus to define the exits
or final outcomes of the system, which will divide the space of initial
conditions into different regions. For the classical system, the three
natural outcomes are given by the three exit trajectories. Any initial point will end up
in one of these three exits, except those that follow an infinite sequence of bounces
against the potential walls and form the repeller set. As explained
in Sec. \ref{sec:analysis_quantum}, those classical exits are closed up by quantum fluctuations.
Nonetheless, our numerical simulations show that, since the fluctuations $B_i/P_Q^2$
are very small, the system takes a very long time to reach those quantum walls,
and therefore, in practice, we can define the three ranges of values
\eqref{exits} as the final outcomes of our quantum system.
In fact, due to the numerical nature of the analysis,
it is necessary to allow an artificial extension of the ranges that define the exits.
Something similar happens under the classical evolution:
since a typical initial point takes a very long time to reach
one of the exits, numerically one can never exactly state that
a given exit has been reached.
Therefore, in order to facilitate the process of escaping and reaching
the outcomes, as done in the classical analysis presented in Ref.~\cite{Cornish_Levin}, a small error
for the exit conditions will be allowed, as we will explain later.

Let us now parameterize the initial values of the variables $({\beta_{+}},{\beta_{-}},{p_{+}},{p_{-}})$ as follows,
\begin{align}\label{initial_conditions}
\begin{split}
{\beta_{+}}|_{\alpha=\alpha_{\rm ini}}&=\frac{1-2v_0(1+u_0)}{2(1+v_0+u_0v_0)}\alpha_0,
\hspace{1cm}
{\beta_-}|_{\alpha=\alpha_{\rm ini}}=-\frac{\alpha_0\sqrt{3}}{2(1+v_0+u_0v_0)},
\\[10pt]
{p_+}|_{\alpha=\alpha_{\rm ini}}&=\frac{\omega_0}{\sqrt{6\pi}}\frac{1-2u_0(1+u_0)}{1+u_0+u_0^2},
\hspace{1cm}
{p_-}|_{\alpha=\alpha_{\rm ini}}=\frac{\omega_0}{\sqrt{2\pi}}\frac{1+2u_0}{1+u_0+u_0^2}
,
\end{split}
\end{align}
with the four real parameters $\alpha_0$, $\omega_0$,
$u_0$, and $v_0$.
In order to explore the boundaries between regions with a different outcome in the
full space of initial data, already for the classical system, one would need to perform an enormous amount
of numerical simulations. Therefore, as done in Ref.~\cite{Cornish_Levin},
we will reduce our study to the cross-section $(u_0,v_0)$ of that space, by
choosing a fixed value of all the initial variables except for $u_0$ and $v_0$.
Furthermore, to compare more efficiently our results with Ref.~\cite{Cornish_Levin},
we will consider the same set of initial conditions for the expectation
values as those considered in that reference for the corresponding classical variables.
More precisely, $\omega_0$ will be fixed to $\omega_0=\frac{1}{3}$, while
$\alpha_0=\alpha_{\rm ini}$ will be the initial time, and its value will be such that the following equation is obeyed,
\begin{equation}
\omega_0=\sqrt{\frac{3\pi}{2}}
H_Q|_{\alpha=\alpha_0},
\end{equation}
with the Hamiltonian $H_Q$ \eqref{quantum_hamiltonian_approx}, which depends on the initial
value of the quantum fluctuations that will be detailed below. Finally, for the numerical analysis,
a $300\times300$ grid in the region defined by $u_0\in[1.34,1.36]$ and $v_0\in[1.3,1.32]$
will be chosen.

Concerning the initial conditions for the quantum variables $(s_1,s_2,p_{s_1},p_{s_2},U_1,U_2)$,
instead of choosing a fixed value for them, and in order to analyze the dimensionality
of the boundaries between regions for different cross-sections of the space of initial data,
we will run a number of numerical evolutions with different sets of values.
In this way, we will be able to check how quantum effects modify the commented cross-section
$(u_0,v_0)$ of the space of initial data, and compute the corresponding value of $\delta$.
In order to fix the precise sets of values in a physically sensible way,
the initial quantum degrees of freedom
will be parameterized in terms of the Planck constant $\hbar$ and the
three numbers $k$, $m$, and $n$ of the order of the unity,
as follows,
\begin{align}
\begin{split}\label{values_simulations}
&s_1|_{\alpha=\alpha_0}=s_2|_{\alpha=\alpha_0}=\sqrt{\frac{k\hbar}{2}},
\\
&p_{s_1}|_{\alpha=\alpha_0}=p_{s_2}|_{\alpha=\alpha_0}= n\sqrt{\frac{\hbar}{2k}},
\\
&U_1|_{\alpha=\alpha_0}=U_2|_{\alpha=\alpha_0}=\frac{\hbar^2}{4}(mk-n^2),
\end{split}
\end{align}
with $n\in\{\pm 1, \pm 1/2,\pm 1/\sqrt{2},0\}$, $k\in\{0.5,
1, 1.1, 1.2,\dots,3\}$, and $m\in\{1,1.1,1.2,\dots,2,3\}$.
Among all these sets, we have chosen a sample of $160$ specific
values ($k$, $m$, $n$),
all satisfying the Heisenberg uncertainty relation \eqref{uncert_heis},
which, within this parameterization, reads $mk-n^2\geq 1$.
  
Moreover, for the numerical simulations the value $\hbar=10^{-6}$ has been considered for the Planck constant,
and the differential equations have been discretized and evolved by making use of a fourth-order adaptive Runge-Kutta method. In particular, at each instant of the evolution,
an initial time step of $d\alpha=10^{-5}$ is considered, which
is iteratively doubled
until the relative error between the corresponding evolutions with a step
$2d\alpha$ and $d\alpha$ exceeds $\rho=10^{-4}$. 
This method allows us to perform long numerical time evolutions during
the Kasner regimes, where the variables do not perform
sudden changes and evolve very smoothly. A decrease of the step size
usually signals the end of the Kasner regime.
In fact, since the momenta $p_\pm$ remain constant during this period,
the numerical algorithm periodically checks
whether the system is in a Kasner regime
by testing if the inequalities $|\dot{p}_{+}|<\rho$ and $|\dot{p}_{-}|<\rho$ are obeyed.
If these conditions are satisfied, the algorithm verifies if the system has reached
(taking into account an error) one of the outcomes \eqref{exits}. More specifically, as mentioned above and following the procedure described in Ref.~\cite{Cornish_Levin}, we allow an error
\begin{align}\label{error_p_+}
\Delta_{\rm num}\left[\frac{p_+}{(p_+^2+p_-^2)^{1/2}}\right]=0.2\left|
\frac{p_-}{(p_+^2+p_-^2)^{1/2}}
\right|,
\end{align}
for the fraction
$p_+/(p_+^2+p_-^2)^{1/2}$, which characterizes the outcomes \eqref{exits}
\footnote{For more information on this error, we refer the reader to App.~\ref{app:error}.}.
Certainly, due to this approximation, some of the initial conditions that constitute the boundary---those that are supposed to never reach a certain outcome---will also escape, but only a very small percentage \cite{Cornish_Levin}.
Therefore, introducing this error minimally changes the division of the space of initial conditions and the subsequent analysis of the uncertainty exponent. 

Depending on which of the three outcomes is reached, a specific color is assigned to that initial condition: blue, red, or green. However, for some specific initial values, the system takes a very long time to reach any of them, even if the algorithm introduces an adaptive time step.
Consequently,
we have introduced a maximum of $2\cdot10^6$ iterations for each initial value.
Those points will not be assigned any color and will be part of the boundaries
between regions. However, in the simulations we have performed, we have checked
that very few initial data reach that maximum of iterations.

Once the distribution of a given cross-section of the space of initial data
is performed, the fraction $f(\varepsilon)$ can be computed for different values of $\varepsilon$
(we have considered around 8--10 values for each sets of data). Finally, according to \eqref{uncertainty_coeff},
simply by plotting $\ln{f(\varepsilon)}$ in terms of $\ln\varepsilon$ and performing a linear regression, the value of the uncertainty exponent $\delta$ can be directly obtained from the slope.

 \begin{figure}
  	\centering
  	\includegraphics[width=0.7\linewidth]{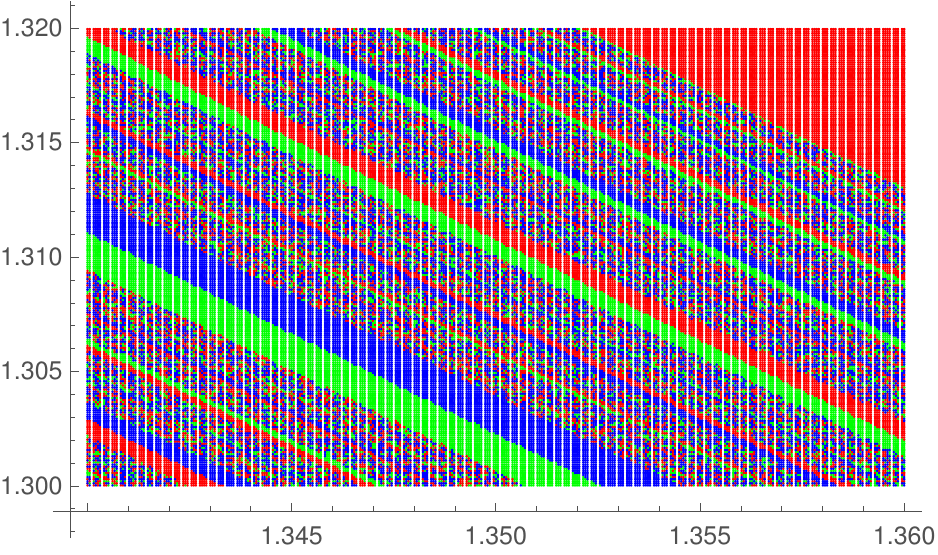}
  	\caption{Division of the space of initial conditions $(u_0,v_0)$ depending on their final outcome, considering a $300\times300$ grid and for the classical model.}
  	\label{fig:grid-compressed}
  \end{figure}
  \begin{figure}
  	\centering
  	\includegraphics[width=0.7\linewidth]{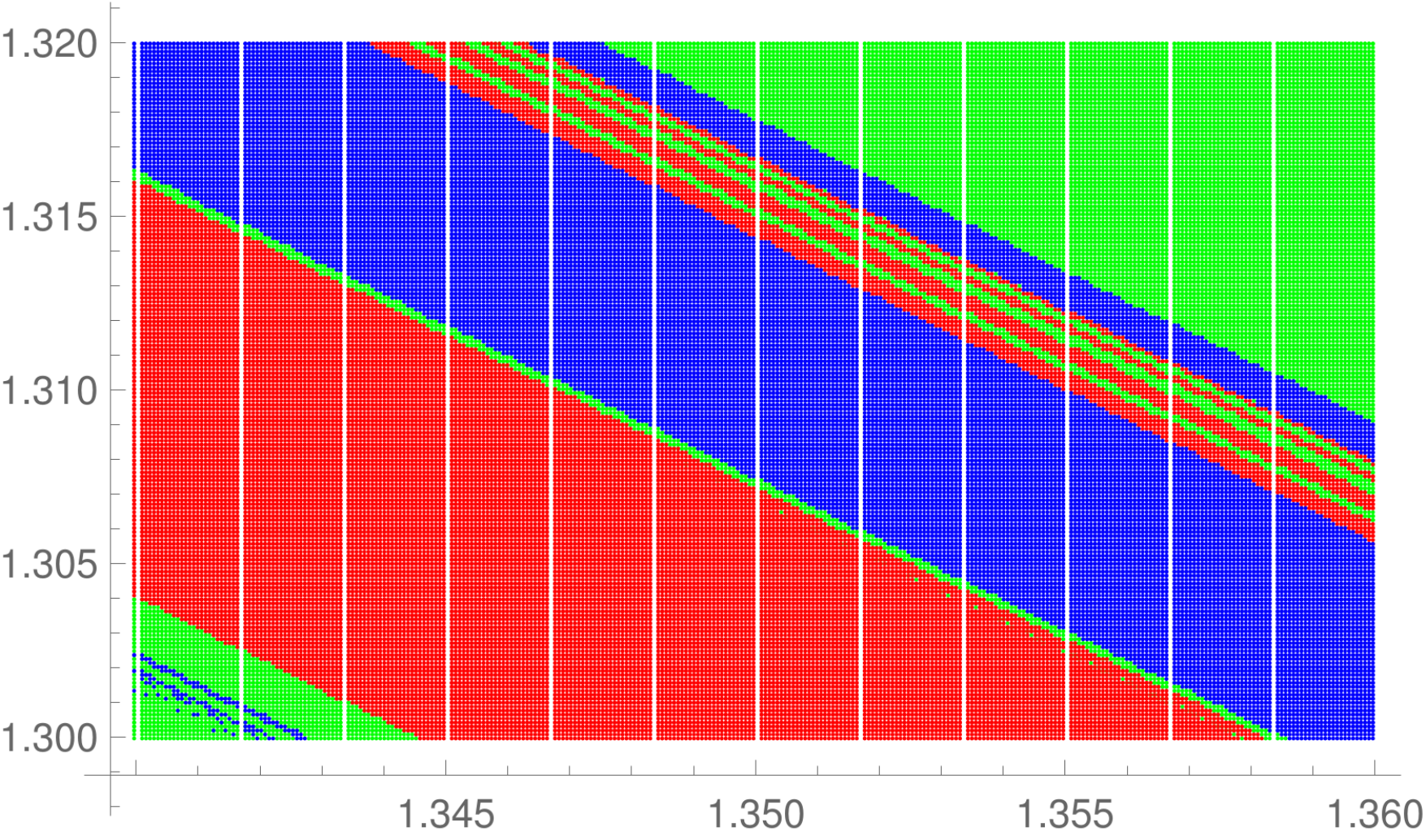}
  	\caption{Division of the space of initial conditions $(u_0,v_0)$ depending on their final outcome, considering a $300\times300$ grid, for the semiclassical model and with $U_1|_{\alpha=\alpha_0}=U_2|_{\alpha=\alpha_0}=\hbar^2/4$, $p_{s_1}=p_{s_2}|_{\alpha=\alpha_0}=-\sqrt{\hbar/2}$
  	and $s_1|_{\alpha=\alpha_0}=s_2|_{\alpha=\alpha_0}=\sqrt{\hbar}/2$.}
  	\label{fig:grid-quantum-compressed}
  \end{figure}
  
 \begin{figure}
 	\centering
 	\begin{subfigure}[h]{0.48\textwidth}
 		\centering
 		\includegraphics[width=\textwidth]{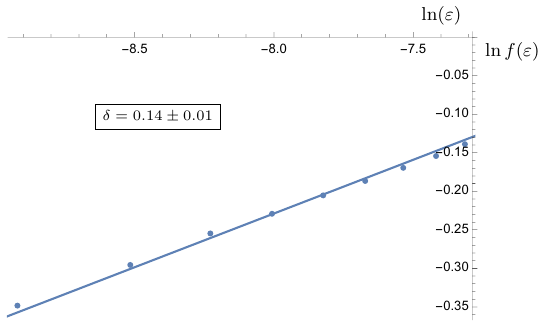}
 	\end{subfigure}
 	\hfill
 	\begin{subfigure}[h]{0.48\textwidth}
 		\centering
 		\includegraphics[width=\textwidth]{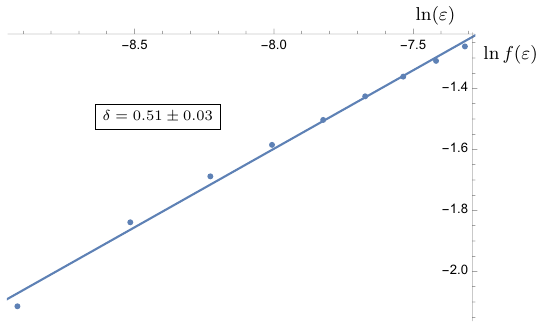}
 	\end{subfigure}
 	\caption{Linear regression of the plot $\ln f(\varepsilon)$ in terms of $\ln\varepsilon$, for the classical model (on the left) and for a quantum state (on the right) with initial $U_1|_{\alpha=\alpha_0}=U_2|_{\alpha=\alpha_0}=\hbar^2/4$,  $p_{s_1}|_{\alpha=\alpha_0}=p_{s_2}|_{\alpha=\alpha_0}=-\sqrt{\hbar/2}$
 		and $s_1|_{\alpha=\alpha_0}=s_2|_{\alpha=\alpha_0}=\sqrt{\hbar}/2$. The value of the uncertainty exponent $\delta$ is given by the slope of the regression.}
 	\label{fig:linear_regression}
 \end{figure}

\subsubsection{Results}
\label{sec:results}

Following the described methodology, in Ref. \cite{Cornish_Levin}, the value $\delta=0.14\pm0.01$ was obtained
for the classical system. For every simulation that we have performed for different
cross-sections of the space of initial data of the quantum model,
a larger value has been obtained, though always $\delta<1$. Specifically, for the sample that we have analyzed we have computed values of $\delta$ in the range $[0.42,0.79]$. 
Therefore, we can conclude that,
while the quantum system is still chaotic, quantum effects increase the value of the uncertainty
coefficient $\delta$, and, thus, considerably reduce its strength.

This can easily be seen, for instance, by comparing Figs. \ref{fig:grid-compressed} and \ref{fig:grid-quantum-compressed}. In the former, 
the distribution of the space of initial data $(u_0,v_0)$ is shown for the
classical model, where each color (red, blue and green) represents one of the
classical exits. In particular, the lack of smoothness of the boundaries conveys its fractal dimension. In the latter, on the contrary, the same space is shown
for the quantum model, with certain particular initial values of the moments
$(s_1,s_2,p_{s_1},p_{s_2},U_1,U_2)$. It is clearly visible that in this case the
boundaries have been smoothed.
This is just a particular example, shown for illustrative purposes, but all the cross-sections
present similar behaviors, i.e., smoother distributions of the different colors in the quantum version.
Furthermore, in order to show quantitatively the reduction of the level of chaos of these two particular examples, in Fig.~\ref{fig:linear_regression} we display the linear regressions performed
in the plot of $\ln f(\varepsilon)$ as a function of $\ln\varepsilon$
for each of them. In the classical case, the value $\delta=0.14$ is recovered,
while the quantum analysis leads to a larger value, $\delta=0.51$ for this particular case.

Moreover, by comparing the simulations for different initial quantum fluctuations,
we have been able to obtain some generic conclusions about the dimensionality of
the boundary for different cross-sections. In particular, we have found that for
initial unsqueezed states, i.e., with the same fluctuation in both conjugate variables
$\Delta(\beta_\pm^2)|_{\alpha=\alpha_0}=\Delta(p_\pm^2)|_{\alpha=\alpha_0}$,
the value of the correlation $\Delta(\beta_{\pm}p_{\pm})|_{\alpha=\alpha_0}$
has a different impact on the parameter $\delta$. Namely,
for states with no initial or positive correlation, $\Delta(\beta_{\pm}p_{\pm})|_{\alpha=\alpha_0}\geq 0$,
the larger the value of the initial fluctuations
$\Delta(\beta_{\pm}^2)|_{\alpha=\alpha_0}=\Delta(p_{\pm}^2)|_{\alpha=\alpha_0}$,
the smaller the value of $\delta$.
However, states with an initial negative correlation $\Delta(\beta_{\pm}p_{\pm})|_{\alpha=\alpha_0}< 0$
show the opposite behavior:
larger values of the initial fluctuations $\Delta(\beta_{\pm}^2)|_{\alpha=\alpha_0}=\Delta(p_{\pm}^2)|_{\alpha=\alpha_0}$, lead to larger values of $\delta$.
Other dependences of $\delta$ on the state properties are more complicated.
For squeezed states, for instance, there is no simple pattern that describes
how the sign of the initial correlation affects the value of $\delta$ as the different fluctuations increase.

\section{Conclusions}\label{sec:discussion}

We have presented a semiclassical model that describes the vacuum Bianchi IX
dynamics, in order to study how the quantum effects modify its classical
chaotic behavior.  The model is semiclassical as it relies on two main
assumptions, namely, the relative fluctuations of the Hamiltonian are supposed
to be small, while the wave function is assumed to keep a Gaussian-like shape
all along evolution. We have implemented this latter condition by a suitable
  parameterization of higher-order moments of the evolving state.  We have
shown that, under such conditions, all the relevant physical information of
the quantum system can be encoded in the Hamiltonian
\eqref{quantum_hamiltonian_approx}. In particular, the classical phase space,
described by the two anisotropy variables and their momenta, is extended by
another two degrees of freedom, which describe the fluctuations of the
anisotropies, and two constants of motion, which contain information about the
saturation of the uncertainty relations.

Concerning the quantum evolution, our numerical resolution of the equations of
motion
\eqref{equations_motion_quantum_beta}--\eqref{equations_motion_quantum_ps} has
shown a similar qualitative picture as the classical dynamics. As it
approaches the singularity, the system follows a succession of kinetically
dominated periods, known as Kasner regimes, and quick transitions, which arise
when the potential term of the Hamiltonian becomes nonnegligible. This
dynamics can be understood as a particle propagating in the anisotropy plane
and bouncing against the potential walls shown in
Fig. \ref{fig:equipotential}.  The main novel feature of the quantum system is
the closing off of the classical exit channels.  If the velocity
  vector of the classical particle is exactly parallel to one of the three
symmetry axes of the potential, its corresponding Kasner regime lasts until
the singularity is reached. For the quantum system, we have analytically shown
that quantum fluctuations widen these classical exit channels, so that the
velocity of the particle does not need to be exactly parallel to the symmetry
axis for it to avoid the classical potential walls.  However, as it evolves,
new quantum terms in the potential progressively become more relevant until,
eventually, they put an end to the corresponding Kasner regime in all the
cases, and prevent the anisotropy parameters from reaching infinity.

Intuitively, instead of the three vertices of the classical potential shown in Fig. \ref{fig:equipotential}, the quantum
system presents three wedges, which are closed off at a finite distance from the origin.
An important consequence of this effect is that the dynamics of the system is now limited
to a finite region of the configuration space. In addition, concerning the possible chaotic behavior
of the system, this picture already suggests that quantum effects might decrease the level of
chaos. While in the classical model all potential walls are convex, and thus defocusing,
which ensures chaos, the closed-off of the classical exits necessarily implies
the presence of certain concave portions of the walls.

In order to perform a more systematic study of chaos of the quantum system,
we note that the present semiclassical model is formulated as a canonical system,
and thus standard dynamical-systems techniques can be applied.
In particular, two key methods have been implemented, which are especially
well suited for systems with reparameterization invariance, such as general relativity.

First, the generalized version of the Misner-Chitre variables
\eqref{misner-chitre}--\eqref{misner-chitre-moments} has been constructed. In
the classical set-up these variables asymptotically define stationary
potential walls on the phase space, which render the dynamically accessible
region compact. In the quantum case, this region is not only compact, but it
also shrinks as the singularity is approached. In fact, in a strict asymptotic
  limit, the region is reduced to hypersurfaces in the configuration space
  that are inaccessible due to the uncertainty relations.  Furthermore, we
have also shown that the dynamics can be mapped to a geodesic motion on a
curved Riemannian manifold, extending in this way the classical billiard
picture.  Making use of the geodesic deviation equation, we have computed the
covariant Lyapunov exponent and found it to be strictly positive, which
characterizes the quantum system to be chaotic.

In a second step, we have numerically computed the dimension of the boundary
between regions with different outcomes in the space of initial data. More
specifically, after numerically computing its evolution, each set of initial data
is characterized by its final state in one of the three possible outcomes of the
system, defined as the three wedges commented above. In
Figs. \ref{fig:grid-compressed} and \ref{fig:grid-quantum-compressed} one can
see a given cross-section of the space of initial data for the classical and
quantum system, respectively, where each color corresponds to a given
outcome. In the classical plot (Fig. \ref{fig:grid-compressed}) the fractal
nature of the boundary between regions with a different outcome is clearly
visible.  However, in the quantum case
(Fig. \ref{fig:grid-quantum-compressed}), the boundaries are smoothed out,
which already signals a decrease of the level of chaos.  This has been
verified by numerically computing the dimension of the mentioned boundary for
a large number of initial data sets. Our results are therefore robust and, in all
the analyzed cases, we have found out that, even if the boundary is still fractal,
its dimension is larger than that of its classical counterpart.  Therefore,
although the quantum system is still chaotic, quantum effects severely reduce
its level.

\section*{Acknowledgments}

DB and SFU thank, respectively, the
Max-Planck-Institut f\"ur Gravitationsphysik
and The Pennsylvania State University
for hospitality while part of this work was done.
SFU was funded by an FPU fellowship and a mobility
scholarship of the Spanish Ministry of Universities.
This work was supported in part by NSF grant PHY-2206591,
the Alexander von Humboldt Foundation,
the Basque Government Grant \mbox{IT1628-22}, and
by the Grant PID2021-123226NB-I00 (funded by MCIN/AEI/10.13039/501100011033 and by “ERDF A way of making Europe”).

\appendix

\section{Computation of the error of the outcomes}\label{app:error}
In this appendix, we provide the relation between the error that we have considered in the algorithm for the outcomes, given by \eqref{error_p_+} in terms of $p_{\pm}$, and the one introduced in Ref.~\cite{Cornish_Levin}, in terms of a single parameter $u\geq 1$. Following the notation of that reference, let us start by defining 
the so-called Kasner exponents:
\begin{align}\label{kasner_exponent}
\begin{split}
&p_1:=\frac{1}{3}\left[
1+\frac{p_+}{(p_+^2+p_-^2)^{1/2}}
+\frac{\sqrt{3}p_-}{(p_+^2+p_-^2)^{1/2}}
\right],
\\
&p_2:=\frac{1}{3}\left[
1+\frac{p_+}{(p_+^2+p_-^2)^{1/2}}
-\frac{\sqrt{3}p_-}{(p_+^2+p_-^2)^{1/2}}
\right],
\\
&p_3:=\frac{1}{3}\left[
1-\frac{2p_+}{(p_+^2+p_-^2)^{1/2}}
\right].
\end{split}
\end{align}
These exponents satisfy the relations $p_1+p_2+p_3=p_1^2+p_2^2+p_3^2=1$, such
that we can parameterize them in terms of a single parameter $u\geq 1$. Specifically,
if one defines $p_{\min}:=\min\{p_1,p_2,p_3\}$, $p_{\max}:=\max\{p_1,p_2,p_3\}$, and $p_{\text{int}}:=\{p_1,p_2,p_3\}-\{p_{\min},p_{\max}\}$, the momenta can be parameterized as follows:
\begin{align}
\label{param_u}
&
p_{\min}=-\frac{u}{1+u+u^2},\quad
p_{\text{int}}=\frac{1+u}{1+u+u^2},\quad
p_{\max}=\frac{u(1+u)}{1+u+u^2}.
\end{align}
From here, one can just compute the value of $u$ directly,
\begin{align}\label{u_def}
u=-\frac{1+p_{\min}+\sqrt{1+2p_{\min}-3p_{\min}^2}}{2p_{\min}}.
\end{align}
The classical exits, $\{p_-<0,p_+=-\frac{p_-}{\sqrt{3}}\}$,
$\{p_->0,p_+=\frac{p_-}{\sqrt{3}}\}$, and $\{p_-=0,p_+\leq 0\}$, correspond to
$p_{\min}=0$, as can be seen from \eqref{kasner_exponent}. Consequently, according to
\eqref{param_u}, they are
characterized by $u\to \infty$. However, if one wants to analyze numerically
whether the system verifies this condition for a given time, it is not
possible to obtain such precision. Therefore, in Ref.~\cite{Cornish_Levin},
they relaxed this requirement and considered $u>8$ as a sufficient condition
for the system to be in an exit. This can be interpreted as widening the exits
shown in Fig.~\ref{fig:equipotential}. Even though this approximation may seem
quite extreme, only about $1\%$ of the uncertain points escape as proved in
Ref.~\cite{Cornish_Levin}, and thus the distribution of the space of
initial data is minimally affected.

Then, if one wants to compute this error in terms of $p_+/(p_+^2+p_-^2)^{1/2}$, which is the ratio
we are using in our set-up to define the exits \eqref{exits}, one just needs to make use of the above definitions \eqref{kasner_exponent} and \eqref{param_u}. First, we obtain that $u>8$ corresponds to the intervals
$p_{\min}\in(-\frac{8}{73},0]$, $p_{\text{int}}\in[0,\frac{9}{73})$, and $p_{\max}\in(\frac{72}{73},1]$
for the different momenta. Thus, we can distinguish six different cases, depending
on which of the Kasner exponents $p_1, p_2$, and $p_3$ defines $p_{\min}$, $p_{\max}$, and $p_{\text{int}}$:
\begin{itemize}
	\item $1^{st}$ case:\quad $p_{1}\in\big(-\frac{8}{73},0\big]$, 
	$p_{2}\in\big[0,\frac{9}{73}\big)$, and $p_{3}\in\big(\frac{72}{73},1\big]$,
	\item $2^{nd}$ case:\quad $p_{1}\in\big(-\frac{8}{73},0\big]$, 
	$p_{2}\in\big(\frac{72}{73},1\big]$, and $p_{3}\in\big[0,\frac{9}{73}\big)$,
	\item 
	$3^{rd}$ case:\quad $p_{1}\in\big[0,\frac{9}{73}\big)$, 
	$p_{2}\in
	\big(-\frac{8}{73},0\big]
	$, and $p_{3}\in\big(\frac{72}{73},1\big]$,
	\item 	$4^{th}$ case:\quad $p_{1}\in
	\big[0,\frac{9}{73}\big)$, 
	$p_{2}\in
	\big(\frac{72}{73},1\big]
	$, and $p_{3}\in\big(-\frac{8}{73},0\big]$,
		\item 	$5^{th}$ case:\quad $p_{1}\in
	\big(\frac{72}{73},1\big]$, 
	$p_{2}\in
	\big[0,\frac{9}{73}\big)
	$, and $p_{3}\in\big(-\frac{8}{73},0\big]$,
	\item 	$6^{th}$ case:\quad $p_{1}\in
	\big(\frac{72}{73},1\big]$, 
	$p_{2}\in
	\big(-\frac{8}{73},0\big]
	$, and $p_{3}\in	\big[0,\frac{9}{73}\big)$.
\end{itemize}
If we invert the relations \eqref{kasner_exponent}, we can write,
\begin{align}\label{kasner_expon_p}
	\frac{p_+}{(p_+^2+p_-^2)^{1/2}}=
	\frac{1}{2}(1-3p_3), \quad
	\quad
	\frac{p_-}{(p_+^2+p_-^2)^{1/2}}=
	\frac{\sqrt{3}}{2}(p_1-p_2).
\end{align}
Consequently, in terms of these ratios, the intervals defined in the six cases above read, 
\begin{itemize}
	\item $1^{st}$ case:\quad $\frac{p_+}{(p_+^2+p_-^2)^{1/2}}\in\big[
	-1,-\frac{143}{146}
\big)$ and $\frac{p_-}{(p_+^2+p_-^2)}\in
\big(-\frac{17\sqrt{3}}{146},0\big]$,
	\item $2^{nd}$ case:\quad $\frac{p_+}{(p_+^2+p_-^2)^{1/2}}\in\big(
	\frac{23}{73},\frac{1}{2}
	\big]$ and $\frac{p_-}{(p_+^2+p_-^2)}\in
	\big(-\frac{40\sqrt{3}}{73},-\frac{\sqrt{3}}{2}\big]$,
	\item 
	$3^{rd}$ case:\quad $\frac{p_+}{(p_+^2+p_-^2)^{1/2}}\in\big[
	-1,-\frac{143}{146}
	\big)$ and $\frac{p_-}{(p_+^2+p_-^2)}\in
	\big[0,\frac{17\sqrt{3}}{146}\big)$,
	\item 	$4^{th}$ case:\quad $\frac{p_+}{(p_+^2+p_-^2)^{1/2}}\in\big[
	\frac{1}{2},
	\frac{97}{146}\big)$ and $\frac{p_-}{(p_+^2+p_-^2)}\in
	\big[-\frac{\sqrt{3}}{2},-\frac{63\sqrt{3}}{146}\big)$,
	\item 	$5^{th}$ case:\quad $\frac{p_+}{(p_+^2+p_-^2)^{1/2}}\in\big[
	\frac{1}{2},
	\frac{97}{146}\big)$ and $\frac{p_-}{(p_+^2+p_-^2)}\in
	\big(\frac{63\sqrt{3}}{146},\frac{\sqrt{3}}{2}\big]$,
	\item 	$6^{th}$ case:\quad $\frac{p_+}{(p_+^2+p_-^2)^{1/2}}\in\big(
	\frac{23}{73},\frac{1}{2}
	\big]$ and $\frac{p_-}{(p_+^2+p_-^2)}\in
	\big[\frac{\sqrt{3}}{2},\frac{40\sqrt{3}}{73}\big)$.
\end{itemize}
As one can note, the intervals for both $\arccos [p_+/(p_+^2+p_-^2)^{1/2}]$ and
$\arcsin [p_-/(p_+^2+p_-^2)^{1/2}]$ have the same length in all the six cases, that is,
\begin{align}\label{delta_arctan}
	\Delta_{\rm num}\left[
	\arccos\left(
	\frac{p_+}{(p_+^2+p_-^2)^{1/2}}
	\right)
	\right]=	\Delta_{\rm num}\left[
	\arcsin\left(
	\frac{p_-}{(p_+^2+p_-^2)^{1/2}}
	\right)
	\right]\approx 0.2.
\end{align}
Finally, from here we can estimate the length of the interval for the ratio $p_+/(p_+^2+p_-^2)^{1/2}$
defined by the condition $u>8$, that is, its corresponding error:
\begin{align}\label{error_p_+_def}
	\Delta_{\rm num}\left[
	\frac{p_+}{(p_+^2+p_-^2)^{1/2}}
	\right]
	\approx \Delta_{\rm num}\left[
	\arccos\left(
	\frac{p_+}{(p_+^2+p_-^2)^{1/2}}
	\right)
	\right]	\left|
	\frac{p_-}{(p_+^2+p_-^2)^{1/2}}
	\right|\approx 0.2
	\left|
	\frac{p_-}{(p_+^2+p_-^2)^{1/2}}
	\right|,
\end{align}
where, in the first equality, we have used that $\Delta_{\rm num} (\arccos(x))\approx \left|\frac{d\arccos(x)}{dx}\right|\Delta_{\rm num} x=\frac{\Delta_{\rm num} x}{\sqrt{1-x^2}}$.


\begin{thebibliography}{9}
	\bibitem{BKL}  V. A. Belinskii, I. M. Khalatnikov, and E. M. Lifschitz, Oscillatory approach to a singular point in the
	relativistic cosmology, Adv. Phys. \textbf{19}, 524 (1970).
	
	\bibitem{Mixmaster_I} C. W. Misner, Mixmaster universe, Phys. Rev. Lett. \textbf{22}, 1071 (1969).
	
	\bibitem{Mixmaster} C. W. Misner, Quantum Cosmology. I, Phys. Rev. \textbf{186}, 1319 (1969).
	
	\bibitem{Berger} B. K. Berger, Numerical approaches to spacetime singularities, Living Rev. Relativity 5, \textbf{1} (2002).
	
	\bibitem{Garfinkle} D. Garfinkle, Numerical simulations of generic singuarities, Phys. Rev. Lett. \textbf{93}, 161101 (2004).
	
	\bibitem{Heinzle} J. M. Heinzle, C. Uggla, and W. C. Lim, Spike oscillations, Phys. Rev. D \textbf{86}, 104049 (2012).
	
	\bibitem{Deterministic_chaos_book}
	\textit{Deterministic Chaos in General Relativity} eds. D Hobill, A. Burd, and A. Coley, (Plenum Press, New York, 1994).
	
	\bibitem{Barrow_1}
	J. D. Barrow and F. Tipler, Analysis of the generic singularity studies by Belinskii, Khalatnikov, and Lifschitz, Phys. Rep.  \textbf{56}, 372 (1979).
	
	\bibitem{Barrow_2}
	J. D. Barrow, Chaos in the Einstein Equations, Phys. Rev. Lett.  \textbf{46}, 963 (1981).
	
	\bibitem{Barrow_3}
	J. D. Barrow, Chaotic behaviour in general relativity, Phys. Rep.  \textbf{85}, 1 (1982).
	
		
	\bibitem{Barrow_4}
	D. F. Chernoff and J. D. Barrow, Chaos in the Mixmaster Universe
	, Phys. Rev. Lett.  \textbf{50}, 134 (1983).
	
	\bibitem{Rugh} S. E. Rugh, Cand. Scient. Thesis, The Niels Bohr Institute, (1990).
	
	\bibitem{Francisco}G. Francisco and G. E. A. Matsas, Qualitative and numerical study of Bianchi IX models, Gen. Rel. Grav. \textbf{20}, 1047 (1988).
	
	\bibitem{Berger_numerical} B. K. Berger, Comments on the computation of Liapunov exponents for the Mixmaster universe, Gen. Rel. Grav. \textbf{23}, 1385 (1991).

	
	\bibitem{Hobill} D. Hobill and D. Bernstein, 
	Classical and Quantum Gravity
	The Mixmaster cosmology as a dynamical system D. Simkins and M. Welge, Class. Quant. Grav. \textbf{8}, 1155 (1991).
	
	\bibitem{Ferraz} K. Ferraz and G. Francisco, Mixmaster numerical behavior and generalizations, Phys. Rev. D \textbf{45}, 1158 (1992).
	
	\bibitem{Szydlowski} M. Szydlowski and A. Krawiec, Average rate of separation of trajectories near the singularity in mixmaster models, Phys. Rev. D \textbf{47}, 5323 (1993).
	
		
	\bibitem{Motter_lyap} 
	A. E. Motter, Relativistic Chaos is Coordinate Invariant,
	Phys. Rev. Lett. \textbf{91}, 231101 (2003).
	
	
	\bibitem{Montani_paper} G.~P.~Imponente and G. Montani, On the Covariance of the Mixmaster Chaoticity,  Phys. Rev. D \textbf{63}, 103501 (2001).
	
	\bibitem{Math_concave} Y. G. Sinai, Dynamical systems with elastic reflections, Russian Mathematical Surveys \textbf{25}, 137 (1970).
	
	    
	\bibitem{Cornish_Levin}
	N.~J.~Cornish and J.~J.~Levin,
	The Mixmaster universe: A Chaotic Farey tale,
	Phys. Rev. D \textbf{55}, 7489 (1997).
		
	\bibitem{Motter_fract}
	A. E. Motter and P. S. Letelier, Mixmaster chaos, Phys. Lett. A \textbf{285}, 127 (2001).
	
	\bibitem{Berger_quantum}
	B. K. Berger, Quantum chaos in the Mixmaster universe, Phys. Rev. D \textbf{39}, 2426 (1989).
	
	\bibitem{Bojowald_Date_1} M. Bojowald, G. Date, and G. M. Hossain, The Bianchi IX model in loop quantum
	cosmology, Class. Quantum Gravity \textbf{21}, 3541 (2004).
	
	\bibitem{Vaulin} A. Kheyfets, W. A. Miller, and R. Vaulin, Quantum geometrodynamics of the Bianchi IX cosmological
	model, Classical Quantum Gravity \textbf{23}, 4333 (2006).
	
	\bibitem{Benini_Montani} R. Benini and G. Montani, Inhomogeneous quantum Mixmaster: from classical toward quantum me-
	chanics, Classical Quantum Gravity \textbf{24}, 387 (2007).
	
	\bibitem{Wilson-Ewing-1}
	 E. Wilson-Ewing, Loop quantum cosmology of Bianchi type IX models, Phys. Rev. D \textbf{82}, 043508 (2010).
	
	\bibitem{Ashtekar} A. Ashtekar, A. Henderson, and D. Sloan, Hamiltonian formulation of the Belinskii-Khalatnikov-Lifshitz
	conjecture, Phys. Rev. D \textbf{83}, 084024 (2011).
	
	\bibitem{Bergeron} H. Bergeron, E. Czuchry, J. P. Gazeau, P. Malkiewicz, and W. Piechocki, Singularity avoidance in a
	quantum model of the Mixmaster universe, Phys. Rev. D \textbf{92}, 124018 (2015).
	
		
	\bibitem{Bae}  J. H. Bae, Mixmaster revisited: wormhole solutions to the Bianchi IX Wheeler–DeWitt equation using
	the Euclidean-signature semi-classical method, Classical Quantum Gravity \textbf{32}, 075006 (2015).
	
\bibitem{Bergeron_2} H. Bergeron, E. Czuchry, J. P. Gazeau, and P. Malkiewicz, Spectral properties of the quantum Mixmaster
	universe, Phys. Rev. D \textbf{96}, 043521 (2017).
	
	\bibitem{Garfinkle_quantum} E. Czuchry, D. Garfinkle, J. R. Klauder, and W. Piechocki, Do spikes persist in a quantum treatment
	of spacetime singularities?, Phys. Rev. D \textbf{95}, 024014 (2017).
	
	\bibitem{Corichi} A. Corichi and E. Montoya, Loop quantum cosmology of Bianchi IX: effective dynamics, Class. Quantum Gravity \textbf{34}, 054001 (2017). 
			
	\bibitem{Kiefer} C. Kiefer, N. Kwidzinski, and W. Piechocki, On the dynamics of the general Bianchi IX spacetime near
	the singularity, Eur. Phys. J. C \textbf{78}, 691 (2018).	
	
	\bibitem{Wilson-Ewing_2} E. Wilson-Ewing, A quantum gravity extension to the Mixmaster dynamics, Class. Quantum Grav. \textbf{36}, 195002 (2019).

	\bibitem{Piechocki} A. G\'o\'zd\'z, W. Piechocki, and G. Plewa, Quantum Belinski-Khalatnikov-Lifshitz scenario, Eur. Phys. J.
	C \textbf{79}, 45 (2019).
	
	\bibitem{Montani_polymer} E. Giovannetti and G. Montani, Polymer representation of the Bianchi IX cosmology in the Misner
	variables, Phys. Rev. D \textbf{100}, 104058 (2019).
	
	 \bibitem{Quantum_Bianchi_IX}
	D. Brizuela and S. F. Uria, Semiclassical study of the mixmaster model: The quantum Kasner map, 
	Phys. Rev. D \textbf{106}, 064051 (2022).
	
	\bibitem{Coley} A. Coley, No chaos in brane world cosmology, Class. Quantum Grav. \textbf{19}, L45 (2002).
		
	\bibitem{Bergeron_3}
	H. Bergeron, E. Czuchry, J. P. Gazeau, P. Malkiewicz, and W. Piechocki, Smooth quantum dynamics
	of the Mixmaster universe, Phys. Rev. D \textbf{92}, 061302 (2015).
		
	\bibitem{Bojowald_Date_2} M. Bojowald, G. Date, Quantum suppression of the generic chaotic behavior
	close to cosmological singularities, Phys. Rev. Lett. \textbf{92}, 071302 (2004).
	
	\bibitem{Montani_chaos_polymer}
	S. Antonioni and G. Montani, Singularity-free and nonchaotic inhomogeneous Mixmaster in polymer representation for the volume of the universe, Phys. Lett. B \textbf{790}, 475 (2019). 
	
	\bibitem{moments}
	M. Bojowald and A. Skirzewski, Effective equations
	of motion for quantum system, Rev. Math. Phys. \textbf{18}, 713 (2006).
	
	\bibitem{moments_2011}
	M. Bojowald, D. Brizuela, H. H. Hernandez, M. J. Koop, and H. A. Morales-Tecotl, High-order quantum
	back-reaction and quantum cosmology with a positive cosmological constant, Phys. Rev. D \textbf{84}, 043514
	(2011).

      \bibitem{JK}
        R. Jackiw and A. Kerman, Time Dependent Variational Principle And The Effective Action,
    Phys. Lett. A \textbf{71}, 158 (1979).

  \bibitem{ABCL}
    F. Arickx, J. Broeckhove, W. Coene,  and P. van Leuven,
    Gaussian Wave-packet Dynamics,
    Int. J. Quant. Chem.: Quant. Chem. Symp. \textbf{20}, 471 (1986).

  \bibitem{P}
    O. Prezhdo, Quantized Hamiltonian Dynamics,
    Theor. Chem. Acc. \textbf{116}, 206 (2006).

  \bibitem{VZ}
    T. Vachaspati and G. Zahariade, A Classical-Quantum Correspondence and Backreaction,
    Phys. Rev. D \textbf{98}, 065002 (2018).
    
	 \bibitem{Martin_moments}
	B. Baytas, M. Bojowald, and S. Crowe, Effective potentials from semiclassical truncation, Phys. Rev. A \textbf{99}, 042114 (2019).
	
	\bibitem{Belinski_book} V. Belinski and M. Henneaux, \textit{The Cosmological Singularit}y (Cambridge University Press, Cambridge,
          England, 2017).

          \bibitem{Description} M. Bojowald, Canonical description of quantum
            dynamics,  J. Phys. A: Math. Theor. \textbf{55}, 504006 (2022).

    \bibitem{Montani_book} G. Montani et al., Classical and Quantum Features of the Mixmaster Singularity, International Journal of Modern Physics A \textbf{23}, 2353 (2008).
    
    \bibitem{Chitre}
    D. M. Chitre, {\it Investigation of Vanishing of a Horizon for Bianchi
Type IX (the Mixmaster) Universe}, PhD thesis, University of Maryland (1972).
    
    \bibitem{mcdonald}
    S.~W. McDonald, C. Grebogi, E. Ott, and J.~A. Yorke,
    Fractal Basin Boundaries,
    Physica D: Nonlinear Phenomena
    \textbf{17}, 125 (1985).
\end{thebibliography}
\end{document}